\documentclass[aps,notitlepage,twocolumn,showpacs,10pt]{revtex4-1}

\usepackage{amsmath,amssymb,bbold,graphicx,xcolor}

\usepackage[pdftex,bookmarks=false,hypertexnames=false,breaklinks=true,colorlinks=true,urlcolor=blue,citecolor=blue,linkcolor=blue]{hyperref}

\newcommand{\taut}{\widetilde\tau}
\newcommand{\bd}{\boldsymbol\delta}
\newcommand{\muAF}{\widetilde\mu}
\newcommand{\DeltaAF}{\widetilde\Delta}

\DeclareMathOperator{\sgn}{sgn}
\DeclareMathOperator{\re}{Re}
\DeclareMathOperator{\im}{Im}

\begin{document}

\title{Spectrum of edge states in the $\boldsymbol{\nu=0}$ quantum Hall phases in graphene}

\author{P. K. Pyatkovskiy}
\author{V. A. Miransky}
\affiliation{Department of Applied Mathematics, Western University, London, Ontario N6A 5B7, Canada}

\begin{abstract}
Edge excitations of the $\nu=0$ quantum Hall state in monolayer graphene are studied within the mean-field theory with different symmetry-breaking terms. The analytical expressions for the continuum (Dirac) model wave functions are obtained for the charge density wave, Kekul\'e distortion, ferromagnetic, and (canted) antiferromagnetic phases. The dispersion equations for each phase and boundary type (zigzag and armchair) are derived, numerically solved, and compared to the results of the corresponding effective tight-binding model. The effect of the next-to-nearest neighbor hopping parameter on the edge state spectrum is studied and revealed to be essential. The criteria for the existence of gapless edge states are established for each phase and edge type. 
\end{abstract}


\maketitle

\section{Introduction}

The unconventional sequence of the integer quantum Hall states at filling factors $\nu=\pm4(n+1/2)$, $n=0,1,2,\dots$\cite{Novoselov2005N,Zhang2005N}, in graphene is a direct experimental manifestation of the Dirac quasiparticles~\cite{Gusynin2005PRL,Peres2006PRB} and the fourfold (spin and valley) degeneracy of Landau levels (LLs) in this system. In high magnetic fields, however, the additional quantum Hall plateaux are observed~\cite{Zhang2006PRL,Jiang2007PRL2,Du2009N,Bolotin2009N}, including the insulating state at the charge neutrality point ($\nu=0$)~\cite{Checkelsky2008PRL,*Checkelsky2009PRB}, which indicates the lifting of the LL degeneracy due to the Coulomb interactions. Several scenarios of the interaction-induced splitting of the lowest Landau level (LLL) leading to the $\nu=0$ state have been proposed, including the charge density wave (CDW)~\cite{Alicea2006PRB,Gusynin2006PRB2,Sheng2007PRL,Herbut2007PRBa}, the Kekul\'e distortion (KD)~\cite{Nomura2009PRL,Hou2010PRB}, the ferromagnetic (F)~\cite{Abanin2006PRL,Abanin2007PRL}, the antiferromagnetic (AF)~\cite{Herbut2007PRBa,Jung2009PRB}, and the canted antiferromagnetic (CAF)~\cite{Herbut2007PRB,Kharitonov2012PRB,Kharitonov2012PRBa,Roy2014} phases. Whereas the bulk energy spectrum is gapped in all these phases, the differences in the edge transport can help to identify the nature of the ground state experimentally. Therefore, it is important to have an accurate theoretical description of the edge state properties for each phase.

Most of the existent studies of edge excitations in the $\nu=0$ quantum Hall state take the simplified approach: the mean-field symmetry breaking potential is assumed to be constant across the sample area~\cite{Abanin2006PRL,Gusynin2008PRB,Gusynin2009PRB,Goerbig2011CRP,Yang2010PRB,Kharitonov2012PRBa}. More rigorous treatment takes into account the modification of the order parameter at the edge~\cite{Fertig2006PRL,*Shimshoni2009PRL,Jung2009PRB,Lado2014PRB,Murthy2014}. Both approaches predict the existence of the current-carrying gapless edge excitations in the F phase~\cite{Abanin2006PRL,Fertig2006PRL,Jung2009PRB}, which rules out this state in the case of a magnetic field perpendicular to the graphene plane, for which the divergent resistance was observed experimentally in Ref.~\cite{Checkelsky2008PRL,*Checkelsky2009PRB} (see also Refs.~\cite{Du2009N,Zhao2012PRL}). The transition from an insulating to a metallic state, which occurs upon tilting the magnetic field~\cite{Young2014N}, supports the scenario of transition between the CAF and F phases~\cite{Kharitonov2012PRBa,Lado2014PRB,note}. The absence of dispersing gapless edge states in the KD phase has been shown for the cases of a particular valley isospin orientation~\cite{Yang2010PRB,Arikawa2011AIPCP} or the simplified confining boundary potential~\cite{Goerbig2011CRP,Goerbig2011RMP}. The edge state spectrum of the armchair graphene ribbon in the CDW and AF phases, obtained numerically by the self-consistent Hartree-Fock calculations, was found to be gapped~\cite{Jung2009PRB}. On the other hand, the analysis done within the continuum (Dirac) model showed that in the case of zigzag edges, the existence of gapless edge states depends on the ratio between the coexisting~\cite{Gorbar2008PRB} chemical-potential-like symmetry breaking term and the corresponding mass gap (assumed to be constant)~\cite{Gusynin2008PRB,Gusynin2009PRB}.

In this paper, we present the systematic study of the edge excitations in the CDW, KD, AF, CAF, and F phases in the cases of ideal zigzag or armchair edges, using the effective Dirac Hamiltonian with constant mean-field symmetry breaking terms. We derive the dispersion equations for the edge states and find the analytic expressions for the corresponding wave functions, taking into account the finite next-to-nearest neighbor (NNN) hopping parameter. 

Besides that, the edge state spectrum is also obtained numerically from the effective tight-binding model for noninteracting electrons where the symmetry-breaking potentials are introduced as the on-site energies and the imaginary NNN hopping parameters. This allows us to calculate the spectrum of edge states between the two valleys (in the case of zigzag edges), where it is not captured by the Dirac model. Within the simplified model, neglecting the modification of the order parameter near the edge, we formulate the most general criteria for the existence of gapless edge excitations for each considered phase and boundary type.

In the case of an armchair ribbon, the spectrum is found to be almost independent of the NNN hopping parameter. We find, in agreement with the previously reported results, that the band gap in the CDW and AF phases is equal to the bulk LLL splitting, and the transition from the CAF to F phase is accompanied by the edge-gap closure. For the KD phase, in general, the spectrum is gapped, however, the edge gap closes at some critical value of the valley isospin angle. This occurs due to the interplay between the bulk order and the effective infinite Kekul\'e mass at the boundary.

In the case of a zigzag ribbon, the finite NNN hopping parameter leads to the deformation of the edge state branch between the two valleys, which makes the energy spectrum gapless (provided the magnitude of the NNN hopping parameter exceeds the LLL splitting) in CDW, AF, CAF, and F phases. The only gapped phase is the KD state, where the edge gap is approximately equal to the half of the bulk LLL splitting. At zero NNN hopping, the spectrum is found to be gapped in the KD phase and gapless in the F phase, whereas for the CDW, AF, and CAF phases the band gap depends on the ratio between the corresponding mass gaps and the chemical-potential-like parameters.

The paper is organized as follows. In Sec.~\ref{secII} we describe the effective continuum mean-field model for the broken symmetry phases, the corresponding tight-binding models, formulate the boundary conditions, and derive the general form of the wave function in the arbitrary phase. The dispersion equations for edge states are written and analyzed numerically for each phase in Sec.~\ref{secIII}. The discussion of the main results is given in Sec.~\ref{secIV}. The expressions for the effective tight-binding Hamiltonians for zigzag and armchair graphene ribbons are provided in the Appendix.

\section{Model and general solution}
\label{secII}

\subsection{Dirac model with broken symmetry}

\begin{figure}
\includegraphics[width=0.5\columnwidth]{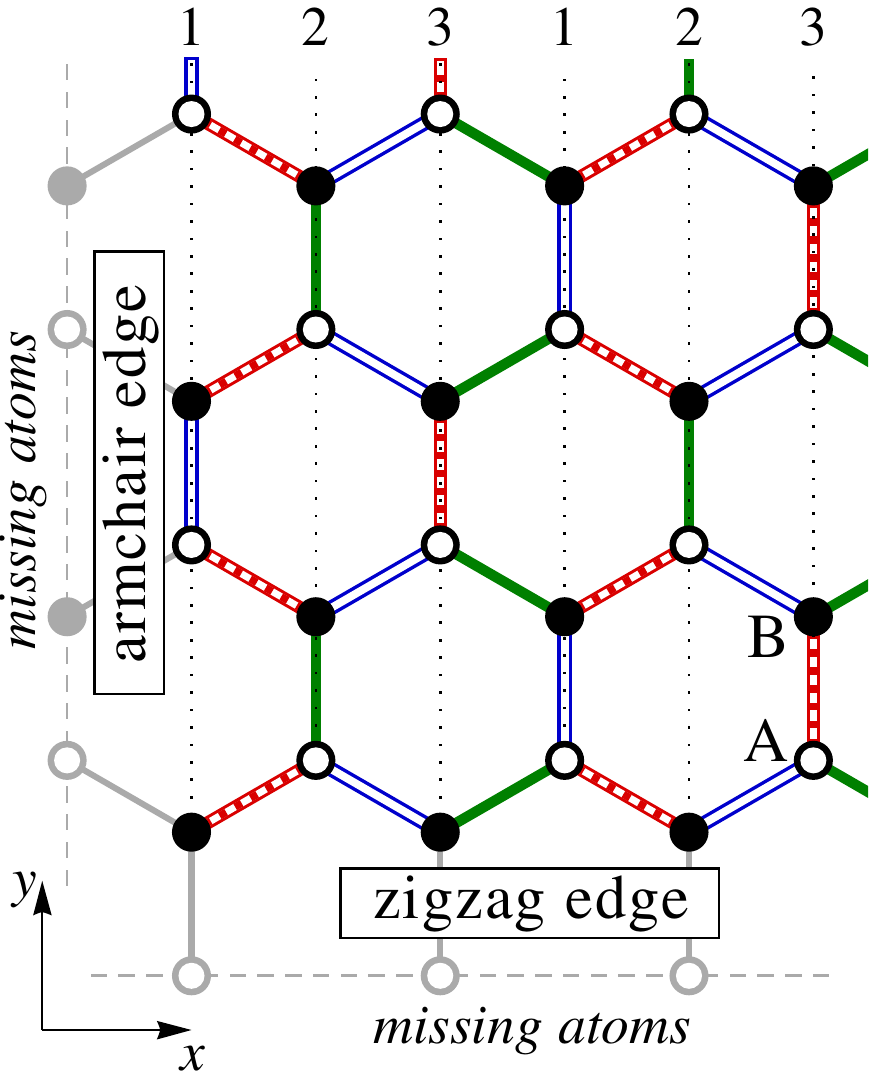}
\caption{Graphene lattice with zigzag and armchair edges. Numbered vertical dotted lines label the three inequivalent atom types within each sublattice in the case of Kekul\'e order (the modulation of the NN hopping parameter is indicated by three different types of lines representing the NN bonds).}
\label{fig:lattice}
\end{figure}

We consider monolayer graphene subject to the external magnetic field $\mathbf B=\nabla\times\mathbf A$ that can be tilted with respect to the $xy$ plane of the two-dimensional lattice (Fig.~\ref{fig:lattice}). The effective mean-field Hamiltonian is $H=H_0+H_1$ with the free $\mathrm U(4)$-symmetric part given by
\begin{equation}
H_0=v_{\rm F}\sigma_0\otimes\taut_0\otimes(\tau_1\hat\pi_x+\tau_2\hat\pi_y).
\label{H0}
\end{equation}
Here $(\hat\pi_x,\hat\pi_y)=-i\hbar\nabla+(e/c)\mathbf A$ is the momentum operator (the electron charge is $-e<0$), $v_{\rm F}=\sqrt3ta/(2\hbar)\simeq10^6$~m/s is the Fermi velocity, $t\simeq3$~eV is the nearest-neighbor (NN) hopping parameter, and $a\simeq0.246$~nm is the lattice constant of graphene. The Pauli matrices $\sigma_i$, $\taut_i$, and $\tau_i$, $i=1,2,3$, act on the spin ($s=\pm$), valley ($K_\pm$), and sublattice ($A$ and $B$) components of the wave function $\Psi=(\Psi^+_{K_+},\Psi^+_{K_-},\Psi^-_{K_+},\Psi^-_{K_-})^T$, respectively, where
\begin{equation}
\Psi^s_{K_+}=
\begin{bmatrix}
\Psi^s_{K_+A} \\
\Psi^s_{K_+B}
\end{bmatrix}
,\qquad
\Psi^s_{K_-}=
\begin{bmatrix}
\Psi^s_{K_-B} \\
-\Psi^s_{K_-A}
\end{bmatrix}
,
\end{equation}
and $\sigma_0$, $\taut_0$, $\tau_0$ are the unit matrices. The basis spin states (the eigenstates of $\sigma_3$) correspond to the direction of the external magnetic field, which does not coincide with the $z$ axis if the field is tilted. The symmetry-breaking part $H_1$ has the general form
\begin{equation}
H_1^{\rm gen}=\sum_{\alpha,\beta=0}^3\sigma_\alpha\otimes\taut_\beta\otimes
(\tau_3\Delta_{\alpha\beta}-\tau_0\mu_{\alpha\beta}).
\label{H_SB_gen}
\end{equation}
It includes the Zeeman splitting term $H_{\rm Z}=\mu_{\rm Z}\sigma_3\otimes\taut_0\otimes\tau_0$ with $\mu_{\rm Z}=\mu_{\rm B}B\approx0.06\,B$[T] meV and the dynamical part, which is mostly generated by the Coulomb interaction. The explicit form of this part depends on a given ground state determined by the interplay between the small lattice-scale asymmetric part of the Coulomb interactions, the Zeeman coupling, and the electron-phonon interactions~\cite{Kharitonov2012PRB}.

In the absence of perpendicular magnetic field, parameters $\mu_{\alpha\beta}$ act like the chemical potentials shifting the Dirac cones for different spins and valleys, whereas the parameters $\Delta_{\alpha\beta}$ result in the (mass) gaps in the bulk spectrum. On the other hand, in the limit of a strong perpendicular magnetic field, within the projection on the LLL, these parameters appear only as a linear combination $\Delta_{\alpha\beta}+\mu_{\alpha\beta}$ and cause the same LLL splitting (but act differently on the higher LLs). However, even in high magnetic fields the edge state spectrum depends on the ratio between $\mu_{\alpha\beta}$ and $\Delta_{\alpha\beta}$ in the case of zigzag edges~\cite{Gusynin2008PRB,Gusynin2009PRB}. The coexistence of these two types of parameters is a general phenomenon and has been explicitly shown for the F and CDW phases in both monolayer~\cite{Gorbar2008PRB} and bilayer~\cite{Gorbar2012PRB} graphene (and also for AF phase in bilayer graphene~\cite{Kharitonov2012PRBb}). We assume here that such a coexistence can also take place for CAF and KD phases. We also assume that the parameters $\mu_{\alpha\beta}$ and $\Delta_{\alpha\beta}$ are determined self-consistently for the infinite graphene sheet and do not vary near the edges of the system. In the following, we consider some specific cases of symmetry-broken phases.

In the mean-field symmetry breaking term
\begin{equation}
H_1^{\rm F}=\sigma_3\otimes\taut_0\otimes(\Delta'\tau_3-\mu'\tau_0)
\label{H_F}
\end{equation}
of the F phase, the finite spin polarization
\begin{equation}
\langle\Psi^\dag\sigma_3\otimes\taut_0\otimes\tau_0\Psi\rangle=
\sum_{s,\xi=\pm}\sum_{X=A,B}s\bigl\langle(\Psi^s_{K_\xi X})^\dag\Psi^s_{K_\xi X}\bigr\rangle
\end{equation}
is described by the enhanced Zeeman splitting $-\mu'\gg\mu_{\rm Z}$. The coexisting order parameter
\begin{equation}
\begin{split}
\langle\Psi^\dag&\sigma_3\otimes\taut_0\otimes\tau_3\Psi\rangle \\
&=\sum_{s,\xi=\pm}s\xi\bigl\langle(\Psi^s_{K_\xi A})^\dag\Psi^s_{K_\xi A}
-(\Psi^s_{K_\xi B})^\dag\Psi^s_{K_\xi B}\bigr\rangle ,
\label{op_delta_so}
\end{split}
\end{equation}
which is dual to $\Delta'$, has the same form as the spin-orbit interaction~\cite{Kane2005PRL}. Note that in Eq.~(\ref{op_delta_so}), the two valleys contribute with the opposite signs; i.e., this order parameter is valley-odd.

The symmetry-breaking part of the Hamiltonian in the CAF phase,
\begin{equation}
H_1^{\rm CAF}=H_1^{\rm F}+H_1^{\rm AF},
\label{H_CAF}
\end{equation}
is characterized by the additional term
\begin{equation}
H_1^{\rm AF}=\sigma_1\otimes\taut_3\otimes(\DeltaAF\tau_3-\muAF\tau_0),
\end{equation}
where we have chosen the $x$ spin axis along the antiferromagnetic vector that can have an arbitrary direction in the plane, perpendicular to the magnetic field~\cite{Herbut2007PRB,Kharitonov2012PRB}. In the purely AF phase (which can exist only in the absence of Zeeman coupling or for the AF vector oriented along, rather then normally to, the magnetic field), the spin density imbalance between the sublattices $A$ and $B$
\begin{equation}
\begin{split}
\langle&\Psi^\dag\sigma_1\otimes\taut_3\otimes\tau_3\Psi\rangle \\
&\;=\sum_{s,s',\xi=\pm}\sigma_1^{ss'}\bigl\langle(\Psi^s_{K_\xi A})^\dag\Psi^{s'}_{K_\xi A}
-(\Psi^s_{K_\xi B})^\dag\Psi^{s'}_{K_\xi B}\bigr\rangle
\end{split}
\end{equation}
connected with $\DeltaAF$ coexists with the valley-odd order parameter
\begin{equation}
\begin{split}
\langle\Psi^\dag\sigma_1&\otimes\taut_3\otimes\tau_0\Psi\rangle \\
&=\sum_{s,s',\xi=\pm}\sum_{X=A,B}\xi\sigma_1^{ss'}\bigl\langle(\Psi^s_{K_\xi X})^\dag\Psi^{s'}_{K_\xi X}\bigr\rangle,
\label{op_mu_af}
\end{split}
\end{equation}
which is dual to $\muAF$.

In the CDW phase,
\begin{equation}
H_1^{\rm CDW}=\sigma_0\otimes\taut_3\otimes(\Delta\tau_3-\mu\tau_0)+H_{\rm Z},
\label{H_CDW}
\end{equation}
the charge imbalance between the sublattices
\begin{equation}
\begin{split}
\langle\Psi^\dag\sigma_0&\otimes\taut_3\otimes\tau_3\Psi\rangle \\
&=\sum_{s,\xi=\pm}\bigl\langle(\Psi^s_{K_\xi A})^\dag\Psi^s_{K_\xi A}
-(\Psi^s_{K_\xi B})^\dag\Psi^s_{K_\xi B}\bigr\rangle
\end{split}
\end{equation}
described by the Dirac mass $\Delta$ coexists with the valley charge imbalance
\begin{equation}
\langle\Psi^\dag\sigma_0\otimes\taut_3\otimes\tau_0\Psi\rangle
=\sum_{s,\xi=\pm}\sum_{X=A,B}\xi\bigl\langle(\Psi^s_{K_\xi X})^\dag\Psi^s_{K_\xi X}\bigr\rangle
\end{equation}
dual to the parameter $\mu$.

The Hamiltonian of the KD phase with the symmetry-breaking term (we use the same variables $\Delta$ and $\mu$ as for the CDW phase)
\begin{equation}
H_1^{\rm KD}=\sigma_0\otimes(\taut_1\cos\theta+\taut_2\sin\theta)\otimes(\Delta\tau_3-\mu\tau_0)+H_{\rm Z}
\label{H_KD}
\end{equation}
is related to its CDW counterpart by the valley isospin rotation
\begin{equation}
H_0+H_1^{\rm KD}=S(H_0+H_1^{\rm CDW})S^\dag
\end{equation}
with
\begin{equation}
S=\frac1{\sqrt2}\sigma_0\otimes(\taut_0+i\taut_1\sin\theta-i\taut_2\cos\theta)\otimes\tau_0,
\label{KD_to_CDW_rotation}
\end{equation}
where the parameter $\theta$ is the valley isospin angle describing the phase of the bond density wave.

\subsection{Representation in the tight-binding model}

The components of the Dirac wave function are related to the tight-binding amplitudes $\psi_s(\mathbf R_X)$ at the atomic sites $\mathbf R_A=\mathbf n\equiv n_1\mathbf a_1+n_2\mathbf a_2$, $\mathbf R_B=\mathbf n+\bd_i$ ($n_1,n_2\in\mathbb Z$) by
\begin{equation}
\psi_s(\mathbf R_X)=\sum_{\xi=\pm}\Psi^s_{K_\xi X}(\mathbf R_X)e^{i\xi\mathbf K\cdot\mathbf R_X},
\label{atomic_psi}
\end{equation}
where
$\pm\mathbf K=(\pm4\pi/(3a),0)$ are the momenta corresponding to $K_\pm$ points, $\mathbf a_1=(a/2,a\sqrt3/2)$, $\mathbf a_2=(a/2,-a\sqrt3/2)$ are the lattice vectors, $\mathbf a_3=-\mathbf a_1-\mathbf a_2$, and the three vectors $\boldsymbol\delta_1=(\mathbf a_1-\mathbf a_2)/3$, $\boldsymbol\delta_2=(\mathbf a_2-\mathbf a_3)/3$, $\boldsymbol\delta_3=(\mathbf a_3-\mathbf a_1)/3$ connect the NN sites (Fig.~\ref{fig:lattice}).

The tight-binding Hamiltonian incorporating only the NN hopping terms is
\begin{equation}
\mathcal H_0=-t\sum_{\mathbf n}\sum_{s=\pm}\sum_{i=1}^3
\bigl(a_{\mathbf ns}^\dag b_{\mathbf n+\boldsymbol\delta_i,s}
+\text{H.c.}\bigr),
\label{H0_TB}
\end{equation}
where $a_{\mathbf R_A,s}$ and $b_{\mathbf R_B,s}$ are Fermi operators corresponding to the atomic orbitals at the sites $\mathbf R_A$ and $\mathbf R_B$. In the continuum limit, it leads to the free Dirac Hamiltonian~(\ref{H0}). The mean-field potentials specific to each phase can be introduced as
\begin{equation}
\begin{split}
&\mathcal H_1^{\rm F}=-\mu'\Omega^+_3-\Delta'\Lambda^-_3, \\
&\mathcal H_1^{\rm AF}=\DeltaAF\Omega^-_1+\muAF\Lambda^+_1, \\
&\mathcal H_1^{\rm CDW}=\Delta\Omega^-_0+\mu\Lambda^+_0+\mathcal H_{\rm Z}, \\
&\mathcal H_{\rm Z}=\mu_{\rm Z}\Omega^+_3,
\end{split}
\end{equation}
where the valley-even symmetry breaking terms are represented by the on-site energies
\begin{equation}
\Omega^\pm_\alpha=\sum_{\mathbf n}\sum_{s,s'=\pm}\sigma_\alpha^{ss'}\bigl(a^\dag_{\mathbf ns}a_{\mathbf ns'}
\pm b_{\mathbf n+\boldsymbol\delta_1,s}^\dag b_{\mathbf n+\boldsymbol\delta_1,s'}\bigr),
\label{Omega}
\end{equation}
and the valley-odd potentials are accounted for by using the imaginary NNN hopping parameters~\cite{Haldane1988PRL,Kane2005PRL,Watanabe2010PRB}:
\begin{equation}
\begin{split}
\Lambda^\pm_\alpha={}&3^{-\frac32}i\sum_{\mathbf n}\sum_{s,s'=\pm}\sum_{i=1}^3
\sigma_\alpha^{ss'}\bigl(a_{\mathbf ns}^\dag a_{\mathbf n+\mathbf a_i,s'} \\
&\pm b_{\mathbf n+\boldsymbol\delta_1,s}^\dag b_{\mathbf n+\boldsymbol\delta_1+\mathbf a_i,s'}-\text{H.c.}\bigr).
\label{Lambda}
\end{split}
\end{equation}
For the KD phase, we use
\begin{align}
\mathcal H_1^{\rm KD}={}&\sum_{\mathbf n}\sum_{s,\kappa=\pm}\sum_{i=1}^3
\Bigl[\frac{\Delta-\kappa\mu}3e^{i\kappa\mathbf K(2\mathbf n+\bd_i)-i\kappa\theta}
a^\dag_{\mathbf ns}b_{\mathbf n+\bd_i,s} \nonumber \\
&+\mbox{H.c.}\Bigr]+\mathcal H_{\rm Z},
\label{KD_term}
\end{align}
where the real and imaginary modulations of the NN hopping $t$ are described by the parameters $\Delta$ and $\mu$, respectively. Note that the hopping parameters are constant along the directions of armchair edges (Fig.~\ref{fig:lattice}). As we will see in Sec.~\ref{secIII}, in a low-energy Dirac model, the abrupt change of the NN hopping parameter from $t$ to zero at the first missing row of bonds at the armchair edge can be viewed as an infinitely large Kekul\'e mass term at the boundary.

Finally, we consider the (real) NNN hopping term:
\begin{equation}
\begin{split}
\mathcal H'&=-t'\sum_{\mathbf n}\sum_{s=\pm}\sum_{i=1}^3
\bigl(a_{\mathbf ns}^\dag a_{\mathbf n+\mathbf a_i,s} \\
&\qquad+b_{\mathbf n+\boldsymbol\delta_1,s}^\dag b_{\mathbf n+\boldsymbol\delta_1+\mathbf a_i,s}+\text{H.c.}\bigr).
\label{HNNN_TB}
\end{split}
\end{equation}
As far as the bulk spectrum is concerned, this term adds a constant $3t'$ to the energy~\cite{Sasaki2006APL} (implicitly subtracted in what follows) and leads to the small LL shifts $\Delta E\sim t'a^2/l^2$~\cite{Kretinin2013PRB}, where $l=\sqrt{\hbar c/(eB_\perp)}$ is the magnetic length.

The magnetic field is introduced in the tight-binding Hamiltonian by the Peierls substitution
\begin{equation}
c_i^\dag c_j\to c_i^\dag c_j\exp\biggl(\frac{ie}{\hbar c}\int_{\mathbf r_i}^{\mathbf r_j}d\mathbf r\cdot \mathbf A\biggr)
\end{equation}
in the hopping terms corresponding to the transitions between the lattice sites $\mathbf r_i$ and $\mathbf r_j$.

\subsection{Boundary conditions}

For a zigzag ribbon $0<y<W$, the tight-binding amplitudes vanish on the first missing rows of atoms (Fig.~\ref{fig:lattice}):
\begin{equation}
\psi_s(\mathbf R_A|_{y=0})=\psi_s(\mathbf R_B|_{y=W})=0.
\end{equation}
This condition uniquely defines the finite difference boundary problem in the case $t'=0$ and translates, according to Eq.~(\ref{atomic_psi}), into the boundary conditions~\cite{Brey2006PRB}
\begin{equation}
\Psi^s_{K_\pm A}(x,0)=\Psi^s_{K_\pm B}(x,W)=0
\end{equation}
for the Dirac model, which also can be written as~\cite{McCann2004JPCM,Akhmerov2008PRB}
\begin{equation}
\begin{split}
&(1+\sigma_0\otimes\taut_3\otimes\tau_3)\Psi(x,0)=0, \\
&(1-\sigma_0\otimes\taut_3\otimes\tau_3)\Psi(x,W)=0.
\label{BC_zigzag}
\end{split}
\end{equation}
In the case $t'\ne0$, the tight-binding equations have to be supplemented with the condition that the amplitudes $\psi_s(\mathbf R_X)$ vanish also on the second missing rows of atoms, and the effective boundary conditions for the Dirac model in this case are~\cite{Wurm2011PRB,Ostaay2011PRB}
\begin{equation}
\begin{split}
\Psi^s_{K_\pm A}(x,0)&=(t'/t)\Psi^s_{K_\pm B}(x,0), \\
\Psi^s_{K_\pm B}(x,W)&=(t'/t)\Psi^s_{K_\pm A}(x,W),
\end{split}
\end{equation}
or, equivalently,
\begin{equation}
\begin{split}
&[1+\sigma_0\otimes\taut_3\otimes(\tau_3\cos\vartheta-\tau_1\sin\vartheta)]\Psi(x,0)=0, \\
&[1-\sigma_0\otimes\taut_3\otimes(\tau_3\cos\vartheta+\tau_1\sin\vartheta)]\Psi(x,W)=0,
\label{BC_zigzag_mod}
\end{split}
\end{equation}
where $\tan(\vartheta/2)=t'/t$.

For the armchair edge at $x=x_0$, the vanishing of the tight-binding amplitudes at the first missing row of atoms (Fig.~\ref{fig:lattice}),
\begin{equation}
\psi_s(\mathbf R_A|_{x=x_0})=\psi_s(\mathbf R_B|_{x=x_0})=0,
\end{equation}
implies, according to Eq.~(\ref{atomic_psi}), the continuum model boundary condition~\cite{Brey2006PRB}
\begin{equation}
\sum_{\xi=\pm}e^{i\xi\theta_0/2}\Psi^s_{K_\xi X}(x_0,y)=0,
\qquad
X=A,B,
\label{BC_armchair}
\end{equation}
which can also be written as~\cite{McCann2004JPCM,Akhmerov2008PRB}
\begin{equation}
[1+\sigma_0\otimes(\taut_2\cos\theta_0-\taut_1\sin\theta_0)\otimes\tau_2]\Psi(x_0,y)=0,
\label{ac_bc}
\end{equation}
where the valley isospin angle $\theta_0=8\pi x_0/(3a)$ depends on the position of the edge.
For a single edge (in the case of a half-plane), the factors $e^{\pm i\theta_0/2}$ in Eq.~(\ref{BC_armchair}) change only the phases of the wave functions in each valley $K_{\xi=\pm}$ and thus are important only in the case of a narrow (compared to the magnetic length) ribbon when the boundary conditions at the opposite edges have to be taken into account simultaneously~\cite{Brey2006PRB,Akhmerov2008PRB} or when the valleys are coupled by the symmetry-breaking term of the bulk Hamiltonian (the KD phase).

\subsection{General solution for the wave function}

In the case of zigzag edges along the $x$ axis, we choose the Landau gauge $(A_x,A_y)=(-B_\perp y,0)$. The wave functions are plane waves in the $x$ direction,
\begin{equation}
\Psi(\mathbf r)=e^{ikx}\Psi(\eta),
\qquad
\eta=y/l-kl,
\end{equation}
and the Dirac equation acquires the form
\begin{equation}
\bigl[-\epsilon_0\sigma_0\otimes\taut_0\otimes
(\tau_+\hat a+\tau_-\hat a^\dag)+H_1^{\rm gen}-E\bigr]\Psi(\eta)=0,
\label{gen_eq}
\end{equation}
where $\tau_\pm=(\tau_1\pm i\tau_2)/2$ are projectors,
$\hat a=2^{-1/2}(\eta+\partial_\eta)$, $\hat a^\dag=2^{-1/2}(\eta-\partial_\eta)$ are the annihilation and creation operators, and $\epsilon_0=\sqrt2\hbar v_F/l$ is the Landau energy scale.
The general solution of this equation is given in terms of the parabolic cylinder functions $U(a,z)$ and $V(a,z)$~\cite{Abramowitz}:
\begin{equation}
\begin{split}
\Psi_{K_\xi}^s(&\eta)=
\sum_i\biggl\{
\begin{bmatrix}
C_{U,1}^{is\xi}U\bigl(\frac12-\lambda_i^{s\xi},\sqrt2\eta\bigr) \\
C_{U,2}^{is\xi}U\bigl(-\frac12-\lambda_i^{s\xi},\sqrt2\eta\bigr)
\end{bmatrix}
\\
&+
\begin{bmatrix}
C_{V,1}^{is\xi}V\bigl(\frac12-\lambda_i^{s\xi},\sqrt2\eta\bigr) \\
C_{V,2}^{is\xi}V\bigl(-\frac12-\lambda_i^{s\xi},\sqrt2\eta\bigr)
\end{bmatrix}
\biggr\}
,\quad
s,\xi=\pm.
\label{gen_sol_zigzag}
\end{split}
\end{equation}
Substituting this solution into Eq.~(\ref{gen_eq}) and using the recurrence relations for the parabolic cylinder functions
\begin{equation}
\begin{split}
\hat aU\bigl(-\tfrac12-\lambda,\sqrt2\eta\bigr)&=\lambda U\bigr(\tfrac12-\lambda,\sqrt2\eta\bigr), \\
\hat a^\dag U\bigl(\tfrac12-\lambda,\sqrt2\eta\bigr)&=U\bigl(-\tfrac12-\lambda,\sqrt2\eta\bigr),  \\
\hat aV\bigl(-\tfrac12-\lambda,\sqrt2\eta\bigr)&=V\bigl(\tfrac12-\lambda,\sqrt2\eta\bigr), \\
\hat a^\dag V\bigl(\tfrac12-\lambda,\sqrt2\eta\bigr)&=\lambda V\bigl(-\tfrac12-\lambda,\sqrt2\eta\bigr)
\end{split}
\end{equation}
leads to the system of algebraic equations. Solving this system for each phase gives the correspondence between the coefficients $C_{U,i}^{is\xi}$, $C_{V,i}^{is\xi}$ and the energy dependence of parameters $\lambda_i^{s\xi}$.

In the following, we will often assume that the ribbon is wide enough ($W\gg l$) so that the bulk LLs are well formed and the states localized near each edge can be considered independently. In this case, one can use the solutions for the half planes $y>0$ and $y<W$ instead of~(\ref{gen_sol_zigzag}).
On a semi-infinite plane $y>0$, the normalizable wave functions contain only the parabolic cylinder functions $U(a,\sqrt2\eta)$ which are bounded at $\eta\to\infty$, and $C_{V,j}^{is\xi}=0$. For the half plane $y<W$, the solution is given by
\begin{equation}
\Psi_{K_\xi}^s(\eta)=
\sum_i
\begin{bmatrix}
-C_{U,1}^{is\xi}U\bigl(\frac12-\lambda_i^{s\xi},-\sqrt2\eta\bigr) \\
C_{U,2}^{is\xi}U\bigl(-\frac12-\lambda_i^{s\xi},-\sqrt2\eta\bigr)
\end{bmatrix}
.\label{gen_sol_other_edge}
\end{equation}
The bulk solutions must be normalizable on an infinite plane and contain only the bounded at $\eta\to\pm\infty$ parabolic cylinder functions
\begin{equation}
U\bigl(-\tfrac12-n,\sqrt2\eta\bigr)=2^{-\frac n2}e^{-\frac{\eta^2}2}H_n(\eta),\quad n=0,1,2,\dots,
\end{equation}
where $H_n(\eta)$ are the Hermite polynomials. This is possible when $\lambda_i^{s\xi}=n$ is a positive integer,
\begin{equation}
\Psi_{K_\xi}^s(\eta)=
e^{-\frac{\eta^2}2}
\begin{bmatrix}
C_{1}^{s\xi}H_{n-1}(\eta) \\
C_{2}^{s\xi}H_n(\eta)
\end{bmatrix}
,
\end{equation}
or when $\lambda_i^{s\xi}=0$ and $C^{is\xi}_{U,1}=0$,
\begin{equation}
\Psi_{K_\xi}^s(\eta)=
e^{-\frac{\eta^2}2}
\begin{bmatrix}
0 \\
C^{s\xi}H_0(\eta)
\end{bmatrix}
.
\end{equation}

For the armchair edges along the $y$ axis, we choose the gauge $(A_x,A_y)=(0,B_\perp x)$. The wave functions are plane waves in the $y$ direction,
\begin{equation}
\Psi(\mathbf r)=e^{iky}\widetilde\Psi(\eta),
\qquad
\eta=x/l+kl,
\end{equation}
and the Dirac equation becomes
\begin{equation}
\bigl[-i\epsilon_0\sigma_0\otimes\taut_0\otimes
(\tau_+\hat a-\tau_-\hat a^\dag)+H_1^{\rm gen}-E\bigr]\widetilde\Psi(\eta)=0.
\end{equation}
Its general solution can be obtained from the solution~(\ref{gen_sol_zigzag}) of Eq.~(\ref{gen_eq}) by the unitary transformation
\begin{equation}
\widetilde\Psi(\eta)=\frac1{\sqrt2}\sigma_0\otimes\taut_0\otimes
(\tau_0+i\tau_3)
\Psi(\eta),
\label{zigzag_to_armchair}
\end{equation}
which does not change the form of $H_1^{\rm gen}$.

\section{Spectra of edge states}
\label{secIII}

\subsection{CDW phase}

\begin{figure}
\includegraphics[width=\columnwidth]{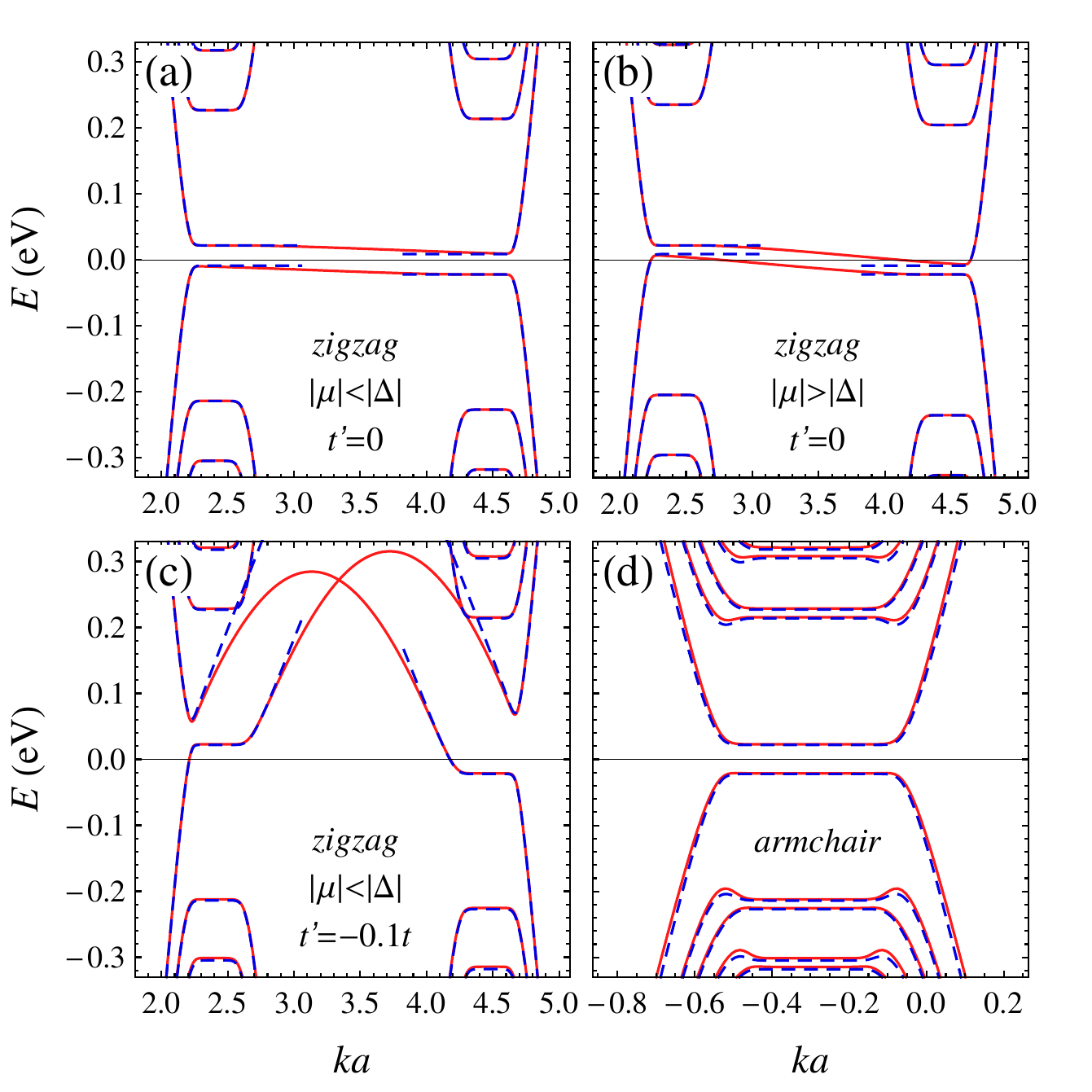}
\caption{Spectrum of graphene ribbons of the width $W=10l$ in perpendicular magnetic field $B_\perp=40$\,T for the CDW phase calculated numerically within the tight-binding (solid line) and Dirac (dashed line) models. Values of parameters used here: (a), (c), (d) $\mu=0.03\epsilon_0$, $\Delta=0.07\epsilon_0$; (b) $\mu=0.07\epsilon_0$, $\Delta=0.03\epsilon_0$. For the armchair ribbon, the NNN hopping $t'=-0.1t$ is taken into account only in the tight-binding calculations. The overall energy shift of $3t'$ is subtracted and the Zeeman splitting is neglected.}
\label{fig:spectra_CDW}
\end{figure}

The symmetry-breaking term~(\ref{H_CDW}) corresponding to the CDW order does not mix different spin ($s=\pm$) and valley ($\xi=\pm$) components that satisfy
\begin{equation}
\bigl[-\epsilon_0(\tau_+\hat a+\tau_-\hat a^\dag)+\xi(\Delta\tau_3-\mu\tau_0)-\mathcal E_s\bigr]\Psi^s_{K_\xi}(\eta)
=0,
\end{equation}
where $\mathcal E_s=E-s\mu_{\rm Z}$. The general solution is given by~\cite{Gusynin2008PRB,Gusynin2009PRB}
\begin{equation}
\begin{split}
\Psi_{K_\xi}^s(\eta)=
{}&
C_{U}^{s\xi}
\begin{bmatrix}
\frac{\mathcal E_s+\xi(\mu+\Delta)}{\epsilon_0}U\bigl(\frac12-\lambda_\xi^s,\sqrt2\eta\bigr) \\
-U\bigl(-\frac12-\lambda_\xi^s,\sqrt2\eta\bigr)
\end{bmatrix}
\\
&+
C_{V}^{s\xi}
\begin{bmatrix}
-V\bigl(\frac12-\lambda_\xi^s,\sqrt2\eta\bigr) \\
\frac{\mathcal E_s+\xi(\mu-\Delta)}{\epsilon_0}V\bigl(-\frac12-\lambda_\xi^s,\sqrt2\eta\bigr)
\end{bmatrix}
,
\label{gen_sol_CDW}
\end{split}
\end{equation}
where
\begin{equation}
\lambda_\xi^s=\bigl[(\mathcal E_s +\xi\mu)^2-\Delta^2\bigr]/\epsilon_0^2.
\label{lambda_CDW}
\end{equation}
The bulk LLs, which correspond to the positive integer values of $\lambda_\xi^s$, are
\begin{equation}
\begin{split}
E_{n\pm}^{s\xi}&=s\mu_{\rm Z}-\xi\mu\pm\sqrt{\epsilon_0^2n+\Delta^2}, \qquad n\ge1, \\
E_0^{s\xi}&=s\mu_{\rm Z}-\xi(\mu+\Delta).
\label{CDW_bulk_spectrum}
\end{split}
\end{equation}
Imposing zigzag boundary conditions~(\ref{BC_zigzag}) on the solution~(\ref{gen_sol_CDW}),
\begin{equation}
\begin{split}
&(\tau_0+\xi\tau_3)\Psi_{K_\xi}^s(-kl)=0, \\
&(\tau_0-\xi\tau_3)\Psi_{K_\xi}^s(W/l-kl)=0,
\label{CDW_zigzag_BC}
\end{split}
\end{equation}
one obtains  $\lambda_\xi^\pm=\lambda_n^\xi(k)$, where $\lambda=\lambda_n^+(k)$ is the $n$th root of the equation~\cite{Gusynin2009PRB}
\begin{equation}
\begin{split}
&\lambda U\bigl(\tfrac12-\lambda,-\sqrt2kl\bigr)V\bigl(-\tfrac12-\lambda,\sqrt2(W/l-kl)\bigr) \\
&-U\bigl(-\tfrac12-\lambda,\sqrt2(W/l-kl)\bigr)V\bigl(\tfrac12-\lambda,-\sqrt2kl\bigr)=0,
\label{CDW_lambda_ribbon_eqs}
\end{split}
\end{equation}
and $\lambda_n^-(k)=\lambda_n^+(W/l^2-k)$. Using Eq.~(\ref{lambda_CDW}), we obtain the energy spectrum
\begin{equation}
E_{n\pm}^{s\xi}(k)=s\mu_{\rm Z}-\xi\mu\pm\sqrt{\Delta^2+\epsilon_0^2\lambda^\xi_n(k)}.
\end{equation}
The lowest solution $\lambda^+_0(k)$ is a monotonically increasing function with $\lambda_0^+(k\to-\infty)\to0$~\cite{Gusynin2009PRB}. This implies that the gap in the energy spectrum is
\begin{equation}
E_{\rm gap}=2(|\Delta|-|\mu|-\mu_{\rm Z})
\end{equation}
[Fig.~\ref{fig:spectra_CDW}(a)]. In the case $E_{\rm gap}<0$, the gapless edge states are present. There is a pair of such states of the same spin that counterpropagate at each edge, and the Dirac model captures only one gapless state from each pair [Fig.~\ref{fig:spectra_CDW}(b)]. The other gapless states are located on the edge state branches connecting the two valleys~\cite{Fujita1996JPSJ,*Nakada1996PRB}, which have a finite dispersion at nonzero $\mu$.

In the case of a finite NNN hopping, applying the boundary conditions~(\ref{BC_zigzag_mod}) to the solution~(\ref{gen_sol_CDW}),
\begin{equation}
\begin{split}
&[\tau_0+\xi(\tau_3\cos\vartheta-\tau_1\sin\vartheta)]\Psi_{K_\xi}^s(-kl)=0,  \\
&[\tau_0-\xi(\tau_3\cos\vartheta+\tau_1\sin\vartheta)]\Psi_{K_\xi}^s(W/l-kl)=0,
\label{CDW_zigzag_BC_y0}
\end{split}
\end{equation}
leads to the dispersion equation
\begin{equation}
\begin{split}
\frac{\mathcal E_s\pm(\mu+\Delta)}{\epsilon_0}&U\bigl(\tfrac12-\lambda_\pm^s,-\sqrt2kl\bigr) \\
\pm\bigl(t'/t\bigr)^{\pm1}&U\bigl(-\tfrac12-\lambda_\pm^s,-\sqrt2kl\bigr)=0.
\label{CDW_disp_eq_zz_mod_0}
\end{split}
\end{equation}
for the edge $y=0$ in the valley $K_\pm$. The corresponding equation for the edge $y=W$ is
\begin{equation}
\begin{split}
\frac{\mathcal E_s\pm(\mu+\Delta)}{\epsilon_0}&U\bigl(\tfrac12-\lambda_\pm^s,\sqrt2(kl-W/l)\bigr) \\
\mp\bigl(t'/t\bigr)^{\mp1}&U\bigl(-\tfrac12-\lambda_\pm^s,\sqrt2(kl-W/l)\bigr)=0.
\label{CDW_disp_eq_zz_mod_W}
\end{split}
\end{equation}
Here we have used $C_{V}^{s\xi}=0$ for the solution on a half plane $y>0$ and took into account Eq.~(\ref{gen_sol_other_edge}) for the solution on a half plane $y<W$.

At finite $t'$, the edge state branches between the two valleys are dispersive. Indeed, within the Dirac model, these edge modes approach the linear asymptotes
\begin{equation}
\begin{split}
E_{y=0}^{s\xi}&\simeq s\mu_{\rm Z}-\xi\mu-\Delta\cos\vartheta+\xi\hbar v_Fk\sin\vartheta, \\
E_{y=W}^{s\xi}&\simeq s\mu_{\rm Z}-\xi\mu+\Delta\cos\vartheta+\xi\hbar v_F(k-W/l^2)\sin\vartheta,
\label{edge_asympt}
\end{split}
\end{equation}
which can be obtained from dispersion equations~(\ref{CDW_disp_eq_zz_mod_0})--(\ref{CDW_disp_eq_zz_mod_W}) by using the asymptotic formula~\cite{Olver1959JRNBS,Pyatkovskiy2013PRB}
\begin{equation}
\frac{U\bigl(\frac12-\lambda,-\sqrt2kl\bigr)}{U\bigl(-\tfrac12-\lambda,-\sqrt2kl\bigr)}\simeq
-\frac{kl+\sqrt{k^2l^2-2\lambda}}{\sqrt2\lambda},
\end{equation}
for $\lambda\gg|kl|\gg1$ (it breaks down at $k>0$, $\lambda\simeq n$, $n\in\mathbb Z$, which corresponds to the avoided crossings with the bulk LLs). Equation~(\ref{edge_asympt}) agrees with the previously obtained dispersion of the corresponding edge modes at zero magnetic field~\cite{Ostaay2011PRB,Tkachov2012PRB}. This result from the Dirac model is a good approximation only in the vicinity of the $K_\pm$ points, and from the tight-binding calculations we see that these edge modes, in fact,  attain their maxima between the two valleys [Fig.~\ref{fig:spectra_CDW}(c)]. In the absence of the symmetry-breaking parameters $\mu$, $\Delta$, and $\mu_{\rm Z}$, the maximum deviation from the LLL energy is equal to $-t'$ and corresponds to the state which is localized exclusively on the outermost row of atoms~\cite{Sasaki2006APL} (it can be easily shown that the effect of experimentally accessible magnetic fields on this state is negligible). Therefore, this edge state branch closes the spectrum gap (if present) provided that $|\Delta/t'|\lesssim1/2$. This condition is expected to be satisfied even for the highest accessible magnetic fields ($B\lesssim50$~T). Indeed, the magnitude of the NNN hopping parameter $t'\simeq-0.3$~eV~\cite{Kretinin2013PRB} exceeds the energy scale $e^2/(\varepsilon_{\rm g}\varepsilon_{\rm s}l)\sim0.01\sqrt{B_\perp[{\rm T}]}$~eV of the Coulomb interactions responsible for the LLL splitting, where the dielectric constants $\varepsilon_{\rm g}=1+\pi e^2/(2\hbar\varepsilon_{\rm s}v_{\rm F})$ and $\varepsilon_{\rm s}$ describe the intrinsic and the substrate-induced screening, respectively~\cite{Goerbig2011RMP}.

In the case of a half plane with the armchair edge at $x=x_0$, the boundary condition~(\ref{ac_bc}) can be rewritten, using Eq.~(\ref{zigzag_to_armchair}), as
\begin{equation}
(1-\taut_2\otimes\tau_1)
\begin{bmatrix}
\Psi^s_{K_+} \\
e^{-i\theta_0}\Psi^s_{K_-}
\end{bmatrix}
_{\eta=kl+x_0/l}=0.
\end{equation}
Substituting the solution~(\ref{gen_sol_CDW}) into this equation with $x_0=0$ gives the dispersion equation~\cite{Gusynin2008PRB}
\begin{equation}
F_1\bigl(\mathcal E_s,\sqrt2kl\bigr)=0,
\label{disp_eqn_armchair_CDW}
\end{equation}
where
\begin{align}
F_1(\mathcal E_s,z)\equiv{}&\frac{\mathcal E_s^2-(\mu+\Delta)^2}{\epsilon_0^2}
U\bigl(\tfrac12-\lambda_+^s,z\bigr)U\bigl(\tfrac12-\lambda_-^s,z\bigr)
\nonumber \\
&-U\bigl(-\tfrac12-\lambda_+^s,z\bigr)U\bigl(-\tfrac12-\lambda_-^s,z\bigr),
\label{F1}
\end{align}
and $\lambda_\xi^s$ are defined in Eq.~(\ref{lambda_CDW}). The dispersion equation for a half plane $x<W$,
\begin{equation}
F_1\bigl(\mathcal E_s,-\sqrt2(kl+W/l)\bigr)=0,
\end{equation}
will be used for the spectrum at the opposite edge of the wide ribbon.
At $\mu=0$, Eq.~(\ref{disp_eqn_armchair_CDW}) reduces to the equation~\cite{Gusynin2008PRB}
\begin{equation}
\lambda\,U^2\bigl(\tfrac12-\lambda,\sqrt2kl\bigr)
-U^2\bigl(-\tfrac12-\lambda,\sqrt2kl\bigr)=0
\label{disp_eqn_armchair_CDW_simpl}
\end{equation}
for $\lambda_+^s=\lambda_-^s=\lambda$, which has the solutions $\lambda=\widetilde\lambda_n(k)$. Taking into account that the lowest solution $\widetilde\lambda_0(k)$ is a monotonic function and $\widetilde\lambda_0(k\to-\infty)\to0$, we see that the spectrum 
\begin{equation}
E_{n\pm}^{s}(k)=s\mu_{\rm Z}\pm\sqrt{\Delta^2+\epsilon_0^2\widetilde\lambda_n(k)}\end{equation}
has a gap of $2(|\Delta|-\mu_{\rm Z})$. In the case of a finite $\mu$, we find numerically that although the lowest energy solution of dispersion equation~(\ref{disp_eqn_armchair_CDW}) can be nonmonotonic, the spectrum gap is still very close to the bulk LLL splitting,
\begin{equation}
E_{\rm gap}\simeq2(|\Delta+\mu|-\mu_{\rm Z}),
\end{equation}
provided that $|\mu|\ll\epsilon_0$ [Fig.~\ref{fig:spectra_CDW}(d)]. The effect of the NNN hopping on the edge gap is also found to be very small if $|t'/t|\ll1$.

\subsection{KD phase}

The symmetry-breaking term~(\ref{H_KD}) of the KD phase mixes the two valleys but leaves the spin components ($s=\pm$) uncoupled:
\begin{equation}
\begin{split}
\bigl[-\epsilon_0\taut_0\otimes(\tau_+\hat a&+\tau_-\hat a^\dag)+(\taut_1\cos\theta+\taut_2\sin\theta) \\
&\otimes(\Delta\tau_3-\mu\tau_0)-\mathcal E_s\bigr]\Psi^{\rm KD}_s(\eta)=0.
\label{KD_eq}
\end{split}
\end{equation}
The general solution $\Psi^{\rm KD}_s(\eta)\equiv[\Psi^{s,\rm KD}_{K_+}(\eta),\Psi^{s,\rm KD}_{K_+}(\eta)]^T$ of the above equation is obtained from the solution $\Psi_s(\eta)\equiv[\Psi^s_{K_+}(\eta),\Psi^s_{K_+}(\eta)]^T$ for the CDW phase~(\ref{gen_sol_CDW}) by the valley isospin rotation~(\ref{KD_to_CDW_rotation}), and the bulk energy spectrum is identical to the spectrum~(\ref{CDW_bulk_spectrum}) of the CDW phase. Imposing the zigzag boundary conditions~(\ref{BC_zigzag}) on the solution of Eq.~(\ref{KD_eq}), one gets the equations
\begin{equation}
\begin{split}
&(1+\taut_3\otimes\tau_3)\Psi^{\rm KD}_s(-kl)=0,  \\
&(1-\taut_3\otimes\tau_3)\Psi^{\rm KD}_s(W/l-kl)=0,
\end{split}
\end{equation}
which are equivalent to the equations
\begin{equation}
\begin{split}
&[1-(\taut_1\cos\theta+\taut_2\sin\theta)\otimes\tau_3]\Psi_s(-kl)=0,  \\
&[1+(\taut_1\cos\theta+\taut_2\sin\theta)\otimes\tau_3]\Psi_s(W/l-kl)=0,
\label{KD_zigzag_BC_y0_mod}
\end{split}
\end{equation}
in terms of the solution~(\ref{gen_sol_CDW}) for the CDW phase. The resulting dispersion equation for a zigzag ribbon,
\begin{equation}
\det
\begin{bmatrix}
Z_+(-kl) & Z_+(W/l-kl) \\
-Z_-(-kl) & Z_-(W/l-kl)
\end{bmatrix}
=0,
\end{equation}
with the $2\times2$ blocks $Z_\pm(\eta)$ defined as
\begin{equation}
\begin{split}
& Z_\pm(\eta) \\
& =\!\begin{bmatrix}
\frac{\mathcal E_s\pm\mu\pm\Delta}{\epsilon_0}U\bigl(\frac12-\lambda^s_\pm,\sqrt2\eta\bigr) &
\pm U\bigl(-\frac12-\lambda^s_\pm,\sqrt2\eta\bigr) \\
V\bigl(\frac12-\lambda^s_\pm,\sqrt2\eta\bigr) &
\!\!\frac{\mu-\Delta\pm\mathcal E_s}{\epsilon_0}V\bigl(-\frac12-\lambda^s_\pm,\sqrt2\eta\bigr)
\end{bmatrix}
\end{split}
\end{equation}
and $\lambda^s_\pm$ given by Eq.~(\ref{lambda_CDW}), is independent of the angle $\theta$.

\begin{figure}
\includegraphics[width=\columnwidth]{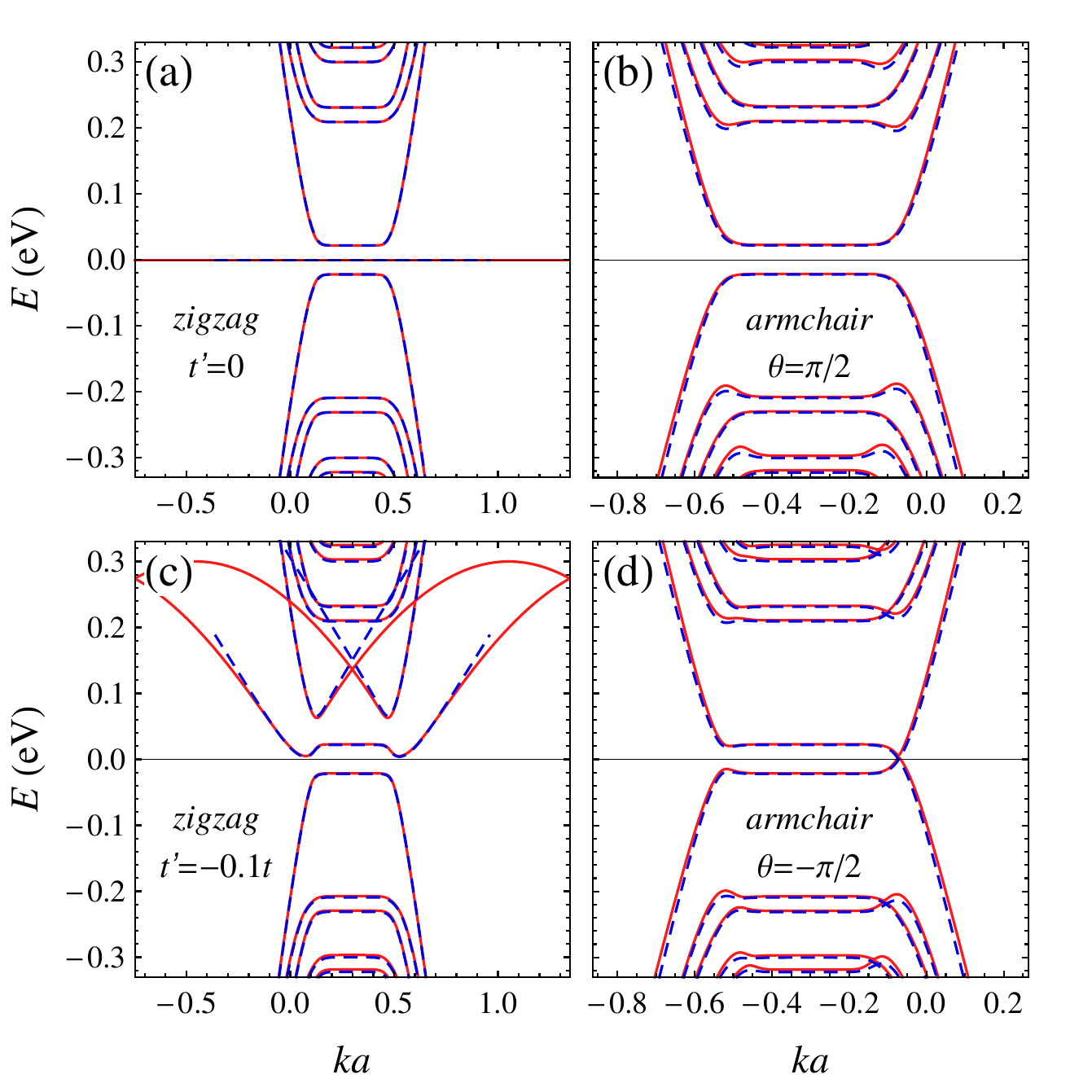}
\caption{Spectrum of graphene ribbons of the width $W=10l$ in perpendicular magnetic field $B_\perp=40$\,T for the KD phase calculated numerically within the tight-binding (solid line) and Dirac (dashed line) models. Values of parameters used here: $\mu=\Delta=0.05\epsilon_0$. For the armchair ribbon ($N=333$), the NNN hopping $t'=-0.1t$ is taken into account only in the tight-binding calculations. The overall energy shift of $3t'$ is subtracted and the Zeeman splitting is neglected.}
\label{fig:spectra_KD}
\end{figure}

The spectrum, shown in Fig.~\ref{fig:spectra_KD}(a), has two dispersionless (in the limit $W\gg l$) edge modes $E=\pm\mu_{\rm Z}$~\cite{Arikawa2011AIPCP,Chang2014JSM}, which lie inside the bulk gap (assuming $2\mu_{\rm Z}<|\mu+\Delta|$).

In the case $t'\ne0$, the modified zigzag boundary conditions~(\ref{BC_zigzag_mod}) applied to the solution at $y=0$,
\begin{equation}
[1-(\taut_1\cos\theta+\taut_2\sin\theta)\otimes(\tau_3\cos\vartheta-\tau_1\sin\vartheta)]\Psi_s(-kl)=0,
\end{equation}
lead to the dispersion equation
\begin{equation}
F_2^{(0)}\bigl(\mathcal E_s,-\sqrt2kl\bigr)-\tan(\vartheta)F_1\bigl(\mathcal E_s,-\sqrt2kl\bigr)=0,
\label{KD_disp_eq_zz_mod}
\end{equation}
where $F_1(\mathcal E_s,z)$ is defined in Eq.~(\ref{F1}) and
\begin{equation}
\begin{split}
F_2^{(n)}(\mathcal E_s,z)\equiv{}&\sum_{\xi=\pm}(\xi)^n\frac{\mathcal E_s+\xi(\mu+\Delta)}{\epsilon_0} \\
&\times U\bigl(\tfrac12-\lambda^s_\xi,z\bigr)U\bigl(-\tfrac12-\lambda^s_{-\xi},z\bigr).
\end{split}
\end{equation}
The dispersion equation for the edge $y=W$ is obtained from Eq.~(\ref{KD_disp_eq_zz_mod}) by replacing $k\to W/l^2-k$. The edge modes are not dispersionless at $t'\ne0$, but in contrast to the CDW phase, the finite edge gap approximately equal to the half of the bulk gap remains even at $|t'|\gg|\mu|,|\Delta|$ [Fig.~\ref{fig:spectra_KD}(c)]. In fact, one can easily check that Eq.~(\ref{KD_disp_eq_zz_mod}) does not have solution $\mathcal E_s=0$ at $t'\ne0$; thus the edge gap is always larger than the half of the bulk gap. In the case $\mu=0$, one can also obtain analytically the ratio between the edge and the bulk gaps:
\begin{equation}
\frac{E_{\rm gap}}{2|\Delta|}=\frac{1+|\sin\vartheta|}2.
\end{equation}

In the case $t'=\mu=0$, one has $\lambda_+^s=\lambda_-^s$ and Eq.~(\ref{KD_disp_eq_zz_mod}) simplifies to
\begin{equation}
\begin{split}
\mathcal E_s U\biggl(-\frac12&-\frac{\mathcal E_s^2-\Delta^2}{\epsilon_0^2},-\sqrt2kl\biggr) \\
&\times U\biggl(\frac12-\frac{\mathcal E_s^2-\Delta^2}{\epsilon_0^2},-\sqrt2kl\biggr)=0.
\end{split}
\end{equation}
In Ref.~\cite{Yang2010PRB}, only the solutions corresponding to the third factor on the left-hand side of the above equation were found.

\begin{figure}
\includegraphics[width=0.72\columnwidth]{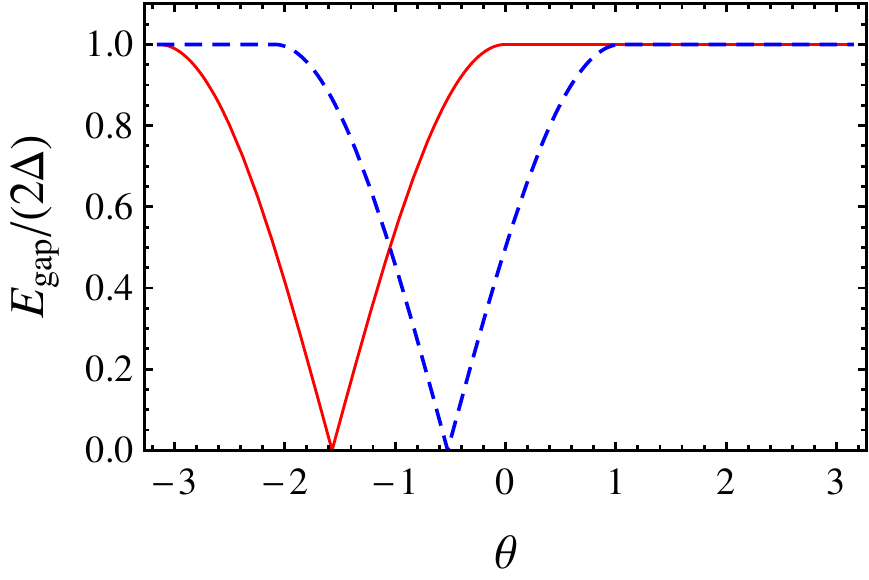}
\caption{Edge gap at $x=0$ (solid line) and $x=W$ (dashed line) in an armchair ribbon as a function of the valley isospin orientation of the KD order parameter. Here $\mu=\mu_{\rm Z}=t'=0$ and $N=3n$, $n\in\mathbb Z$. For $N=3n\pm1$, the dashed curve is shifted horizontally with $\theta\to\theta\mp2\pi/3$.}
\label{fig:gap_KD}
\end{figure}

For the armchair edge $x=x_0$, the boundary condition~(\ref{ac_bc}) can be written as
\begin{equation}
\bigl[1+(\taut_2\cos\theta_0-\taut_1\sin\theta_0)\otimes\tau_2\bigr]\widetilde\Psi^{\rm KD}_s(kl+x_0/l)=0,
\end{equation}
or, using Eqs.~(\ref{zigzag_to_armchair}) and~(\ref{KD_to_CDW_rotation}),
\begin{equation}
\begin{split}
\Bigl\{1+\bigl[&(\taut_1\sin\theta-\taut_2\cos\theta)\cos(\theta-\theta_0) \\
&-\taut_3\sin(\theta-\theta_0)\bigr]\otimes\tau_1\Bigr\}\Psi_s(kl+x_0/l)=0,
\label{BC_KD_W}
\end{split}
\end{equation}
in terms of the solutions for the CDW phase. For the half plane $x>0$, we use the solution~(\ref{gen_sol_CDW}) with $C_V^{s\xi}=0$, which leads to the dispersion equation
\begin{equation}
F_1\bigl(\mathcal E_s,\sqrt2kl\bigr)-\sin(\theta)F_2^{(1)}\bigl(\mathcal E_s,\sqrt2kl\bigr)=0.
\label{KD_disp_eq_hp_0}
\end{equation}
In the case $\sin\theta=0$, it simplifies to the corresponding dispersion equation~(\ref{disp_eqn_armchair_CDW}) for the CDW phase, in particular, for $\mu=0$ it reduces to Eq.~(\ref{disp_eqn_armchair_CDW_simpl})~\cite{Yang2010PRB}. At $\theta=\pm\pi/2$, Eq.~(\ref{KD_disp_eq_hp_0}) can be factorized into two equations:
\begin{equation}
\frac{\xi\mathcal E_s+\mu+\Delta}{\varepsilon_0}U\bigl(\tfrac12-\lambda_\xi^s,\sqrt2kl\bigr)
\pm U\bigl(-\tfrac12-\lambda_\xi^s,\sqrt2kl\bigr)=0,
\label{KD_disp_eq_factorized}
\end{equation}
where $\xi=\pm$ correspond to the eigenstates of the valley isospin matrix $\taut_2$. In particular, when $\theta=\theta_{\rm cr}$ with the critical angle
\begin{equation}
\theta_{\rm cr}=-\sgn(\mu+\Delta)\frac\pi2,
\end{equation}
Eq.~(\ref{KD_disp_eq_factorized}) has a solution $\mathcal E_s=0$ for each $\xi=\pm$ and the spectrum is gapless [Fig.~\ref{fig:spectra_KD}(d)]. In the case $\mu=0$, one can obtain analytically from Eq.~(\ref{KD_disp_eq_hp_0}) the $\theta$ dependence of the ratio between the edge and the bulk gaps (Fig.~\ref{fig:gap_KD}):
\begin{equation}
\frac{E_{\rm gap}}{2|\Delta|}=
\left\{
\begin{array}{cl}
|\cos\theta|, &\quad \Delta\sin\theta<0, \\
1, &\quad \Delta\sin\theta>0.
\end{array}
\right.
\end{equation}
Qualitatively similar behavior (with $\Delta$ replaced by $\Delta+\mu$) is observed numerically for $\mu\ne0$. At nonzero $\mu_{\rm Z}$, the spectrum is gapless for a finite range of $\theta$, namely, when $|\theta-\theta_{\rm cr}|\lesssim \mu_{\rm Z}/|\Delta+\mu|$.

For the edge $x=W$, applying the boundary condition~(\ref{BC_KD_W}) to the corresponding solution for a half plane $x<W$ [see Eq.~(\ref{gen_sol_other_edge})] yields Eq.~(\ref{KD_disp_eq_hp_0}) with $k\to-k-W/l^2$, $\theta\to\theta_W-\theta$, where
\begin{equation}
\theta_W=\frac{8\pi W}{3a}=\frac{4\pi(N+1)}3,
\end{equation}
and the dispersion equation is written as
\begin{equation}
\begin{split}
F_1&\bigl(\mathcal E_s,-\sqrt2(kl+W/l)\bigr) \\
&+\sin\bigl(\theta+\tfrac{2\pi m}3\bigr)F_2^{(1)}\bigl(\mathcal E_s,-\sqrt2(kl+W/l)\bigr)=0.
\label{KD_disp_eq_hp_W}
\end{split}
\end{equation}
Here $m=0,\pm1$ depends on the number of atoms $N$ across the ribbon, $N+1=3n+m$, $n\in\mathbb Z$. For the edge $x=W$, the critical angle at which Eq.~(\ref{KD_disp_eq_hp_W}) has solution $\mathcal E_s=0$ is $\theta'_{\rm cr}=-\theta_{\rm cr}-2\pi m/3$. This implies that the spectrum cannot be gapless at both edges simultaneously (Fig.~\ref{fig:gap_KD}).

Edge-gap closing at a critical valley isospin angle of the KD order parameter was recently pointed out in a tight-binding study of Ref.~\cite{Chang2014JSM} at $B=0$. In the case of a strong magnetic field, this phenomenon can be simply understood by noticing that the armchair boundary condition~(\ref{ac_bc}) is equivalent to the infinite Kekul\'e mass boundary term (rather than the infinite Dirac mass boundary condition commonly used in graphene~\cite{Tworzydlo2006PRL,Furst2009NJP,Stauber2009NJP})
\begin{equation}
\begin{split}
V_{\rm conf}(x)={}&n_xM(x)\sigma_0\otimes(\taut_2\cos\theta_0-\taut_1\sin\theta_0)\otimes\tau_3,
\label{ac_bc_BM_KD}
\end{split}
\end{equation}
which confines the motion of electrons to the region $n_x(x-x_0)<0$, where $M(x)=M_0\Theta(n_x(x-x_0))$, $n_x=\pm1$ is the $x$ component of the outward unit vector normal to the boundary, $\Theta(x)$ is the Heaviside step function, and $M_0\to+\infty$. Indeed, for a two-component spinor the confining Berry-Mondragon mass term $M(x)\tau_3$ implies the boundary condition $(1-n_x\tau_2)\Psi_{K_\xi}^s(x_0,y)=0$~\cite{Berry1987PRSLA}, from which the armchair boundary condition~(\ref{ac_bc}) is obtained by adding the valley matrix structure $\taut'\equiv\taut_2\cos\theta_0-\taut_1\sin\theta_0$. In the absence of valley symmetry breaking in the bulk, $V_{\rm conf}(x)$ produces the edge splitting of LLs that correspond to the different $\taut'$ eigenstates~\cite{Fertig2006PRL,Tworzydlo2007PRB,Kharitonov2012PRBa}. The spatially homogeneous LLL splitting of the same valley components is caused by the KD symmetry-breaking term~(\ref{H_KD}) with $\theta=\theta_0\pm\pi/2$. In particular, when $\theta=\theta_0\pm n_x\theta_{\rm cr}$, the constant bulk and growing near the edge boundary contributions have the opposite signs and cancel each other at some distance from the edge; i.e., the gap closes.

Note that in the case of a smooth confining Dirac mass potential
$\widetilde V_{\rm conf}(y)=V(y)\sigma_0\otimes\taut_3\otimes\tau_3$,
the edge state spectrum of the KD phase was found to be gapped~\cite{Goerbig2011RMP,Goerbig2011CRP}. For this type of boundary one should expect, by the same argument, that the spectrum is gapless in the CDW phase with the appropriate sign of $\Delta+\mu$. Indeed, for the abruptly changing at $y=0$ potential $V(y)=M_0\Theta(-y)$, which is equivalent to imposing the boundary conditions $(1-\xi\tau_1)\Psi_{K_\xi}^s(-kl)=0$ on the solution~(\ref{CDW_zigzag_BC_y0}) with $C_V^{s\xi}=0$, one obtains in the $K_\xi$ valley the same dispersion equation~(\ref{KD_disp_eq_factorized}) with the upper sign.

\subsection{AF, CAF, and F phases}

While we consider the Hamiltonian~(\ref{H_CAF}) of the CAF phase in general, the F and AF phases are treated as the special cases with $\DeltaAF=\muAF=0$ and $\Delta'=\mu'=0$, respectively. The valley components $\Psi_{K_\xi}(\eta)=[\Psi^+_{K_\xi}(\eta),\Psi^-_{K_\xi}(\eta)]^T$ decouple and the energy eigenvalue equation for each valley ($\xi=\pm$) reads
\begin{equation}
\begin{split}
\bigl[-\epsilon_0\sigma_0\otimes(\tau_+\hat a&+\tau_-\hat a^\dag)
+\sigma_3\otimes(\Delta'\tau_3-\mu'\tau_0) \\
+\xi\sigma_1&\otimes(\DeltaAF\tau_3-\muAF\tau_0)-E\bigr]\Psi_{K_\xi}(\eta)=0.
\end{split}
\end{equation}
The general solution is given by
\begin{equation}
\begin{split}
\Psi_{K_\xi}(\eta)=&\sum_{\kappa=\pm}
\Biggl\{
C_U^{\kappa\xi}
\begin{bmatrix}
a_\kappa^+U\bigl(\frac12-\lambda_\kappa,\sqrt2\eta\bigr)
\\
-b_\kappa U\bigl(-\frac12-\lambda_\kappa,\sqrt2\eta\bigr)
\\
\xi c_\kappa^+U\bigl(\frac12-\lambda_\kappa,\sqrt2\eta\bigr)
\\
\xi d^+U\bigl(-\frac12-\lambda_\kappa,\sqrt2\eta\bigr)
\end{bmatrix}
\\
&+C_V^{\kappa\xi}
\begin{bmatrix}
b_\kappa V\bigl(\frac12-\lambda_\kappa,\sqrt2\eta\bigr)
\\
-a_\kappa^-V\bigl(-\frac12-\lambda_\kappa,\sqrt2\eta\bigr)
\\
\xi d^-V\bigl(\frac12-\lambda_\kappa,\sqrt2\eta\bigr)
\\
\xi c_\kappa^-V\bigl(-\frac12-\lambda_\kappa,\sqrt2\eta\bigr)
\end{bmatrix}
\Biggr\},
\label{gen_sol_CAF}
\end{split}
\end{equation}
where we introduced
\begin{equation}
\begin{split}
a_\kappa^\pm&=(E\pm\Delta'-\mu')g^\pm-(E\pm\Delta'+\mu')\epsilon_0^2\lambda_\kappa,
\\
b_\kappa&=\epsilon_0\bigl[(E-\mu')^2-\Delta'^2-\DeltaAF^2+\muAF^2-\epsilon_0^2\lambda_\kappa\bigr],
\\
c_\kappa^\pm&=(\DeltaAF\mp\muAF)g^\pm-(\DeltaAF\pm\muAF)\epsilon_0^2\lambda_\kappa,
\\
d^\pm&=2\epsilon_0\bigl[\muAF(\Delta'\pm E)-\mu'\DeltaAF\bigr],
\\
g^\pm&=E^2-(\mu'\pm\Delta')^2-(\muAF\pm\DeltaAF)^2,
\end{split}
\end{equation}
and
\begin{equation}
\begin{split}
\lambda_\pm={}&\frac1{\epsilon_0^2}\biggl(E^2+\mu'^2+\muAF^2-\Delta'^2-\DeltaAF^2 \\
&\pm2\sqrt{E^2(\mu'^2+\muAF^2)-(\mu'\DeltaAF-\muAF\Delta')^2}\biggr).
\label{lambda_CAF}
\end{split}
\end{equation}

\begin{figure*}
\includegraphics[width=\textwidth]{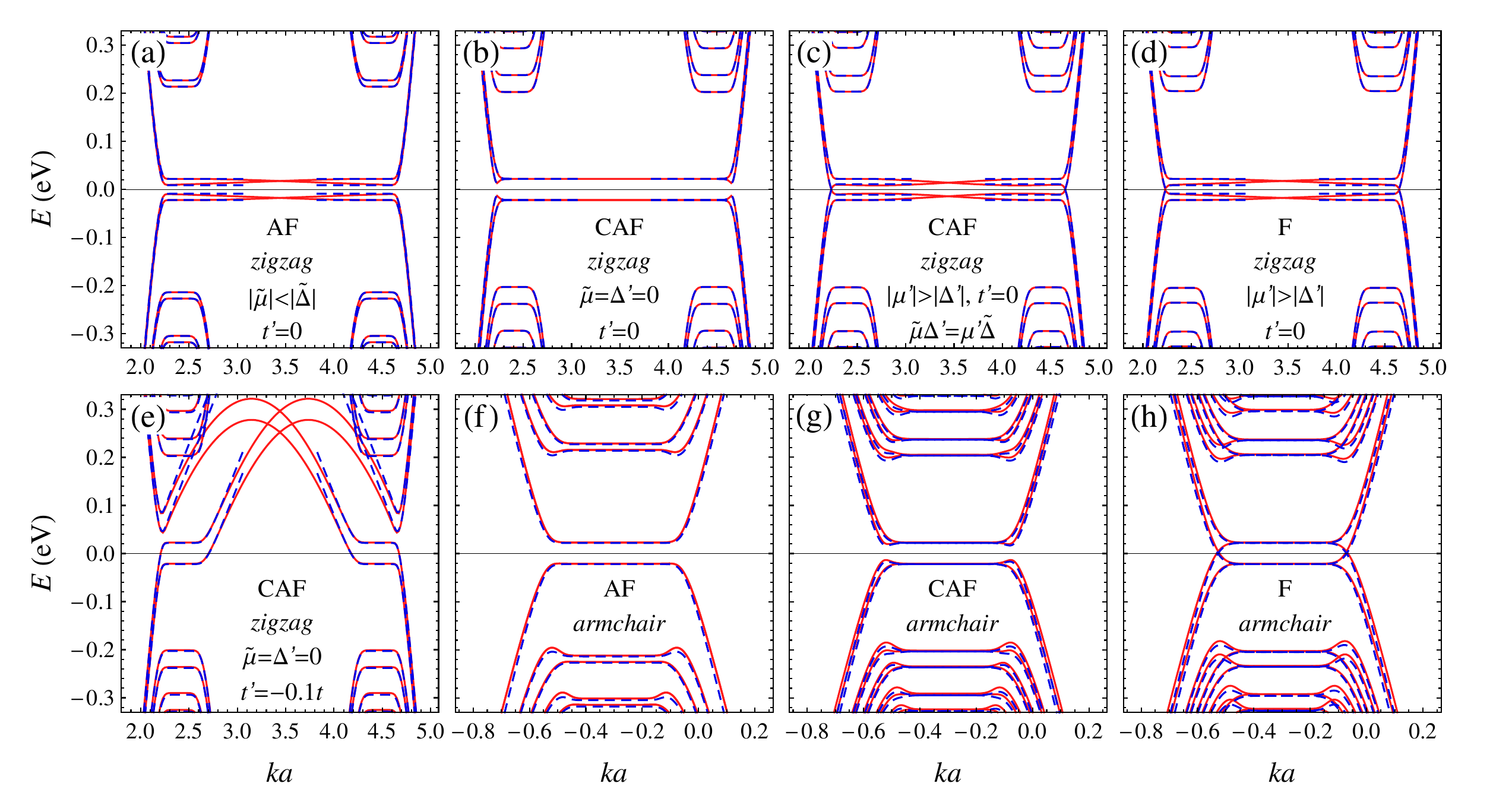}
\caption{
Spectrum of graphene ribbons of the width $W=10l$ in perpendicular magnetic field $B_\perp=40$\,T for the AF, CAF, and F phases calculated numerically within the tight-binding (solid line) and Dirac (dashed line) models.  Values of parameters used here: (a), (f) $\mu'=\Delta'=0$, $\muAF=0.03\epsilon_0$, $\DeltaAF=0.07\epsilon_0$; (b), (e) $\muAF=\Delta'=0$, $\mu'=-0.08\epsilon_0$, $\DeltaAF=0.06\epsilon_0$; (c), (g) $\mu'=-0.045\epsilon_0$, $\Delta'=-0.015\epsilon_0$, $\muAF=0.06\epsilon_0$, $\DeltaAF=0.02\epsilon_0$; (d), (h) $\muAF=\DeltaAF=0$, $\mu'=-0.07\epsilon_0$, $\Delta'=-0.03\epsilon_0$. For the armchair ribbon, the NNN hopping $t'=-0.1t$ is taken into account only in the tight-binding calculations. The overall energy shift of $3t'$ is subtracted.}
\label{fig:spectra_CAF}
\end{figure*}

The bulk LLs corresponding to the positive integer values of $\lambda_{\kappa}$ are
\begin{equation}
\begin{split}
E_{n\pm}^\kappa&=\pm
\biggl[\beta^2+\Bigl(\alpha+\kappa\sqrt{\gamma^2+\epsilon_0^2n}\Bigr)^2\biggr]^{1/2}, \\
&\qquad\kappa=\pm,\quad n\ge1, \\
E_{0\pm}&=\pm\sqrt{(\mu'+\Delta')^2+(\muAF+\DeltaAF)^2},
\label{LLL_CAF}
\end{split}
\end{equation}
where
\begin{equation}
\begin{split}
\alpha&=\sqrt{\mu'^2+\muAF^2}, \\
\beta&=(\mu'\DeltaAF-\muAF\Delta')/\alpha, \\
\gamma&=(\mu'\Delta'+\muAF\DeltaAF)/\alpha.
\end{split}
\end{equation}
Note that the ferromagnetic parameter $\mu'$ includes the bare Zeeman splitting $\mu_{\rm Z}$. In the special case when the valley-odd potentials are absent ($\muAF=\Delta'=0$), the bulk spectrum~(\ref{LLL_CAF}) reduces to~\cite{Herbut2007PRB,Roy2014,Kharitonov2012PRBa}
\begin{equation}
\begin{split}
E_{n\pm}^\kappa&=\pm\sqrt{\DeltaAF^2+\bigl(|\mu'|+\kappa\epsilon_0\sqrt n\bigr)^2},\quad n\ge1, \\
E_{0\pm}&=\pm\sqrt{\DeltaAF^2+\mu'^2}.
\end{split}
\end{equation}

Imposing the zigzag boundary conditions~(\ref{BC_zigzag}) at the two edges of the ribbon,
\begin{equation}
\begin{split}
&\sigma_0\otimes(\tau_0+\xi\tau_3)\Psi_{K_\xi}^s(-kl)=0, \\
&\sigma_0\otimes(\tau_0-\xi\tau_3)\Psi_{K_\xi}^s(W/l-kl)=0,
\end{split}
\end{equation}
one arrives at two identical equations~(\ref{CDW_lambda_ribbon_eqs}) for $\lambda=\lambda_\pm$ in the $K_+$ valley and the corresponding equations for the CDW phase in the $K_-$ valley. Therefore, in the $K_\xi$ valley one has $\lambda_\pm=\lambda^n_\xi(k)$. As follows from Eq.~(\ref{lambda_CAF}), this implies the energy spectrum
\begin{equation}
\begin{split}
E^{\xi\kappa}_{n\pm}(k)&
=\pm\sqrt{\beta^2+\Bigl(\alpha+\kappa\sqrt{\gamma^2+\epsilon_0^2\lambda_n^\xi(k)}\Bigr)^2}, \\
&\qquad \kappa=\pm, \quad n=0,1,2,\dots.
\label{spectr_caf_zz}
\end{split}
\end{equation}
Taking into account that the lowest solution $\lambda_0^\xi(k)$ changes continuously from $0$ to $+\infty$, we see that the lowest energy branch is monotonic if $|\gamma|>\alpha$ or has an extremum otherwise. Thus, the spectrum gap in the Dirac model is equal to
\begin{equation}
E_{\rm gap}=2\sqrt{\beta^2+\Theta(|\gamma|-\alpha)(|\gamma|-\alpha)^2}.
\end{equation}

In the case $\mu'\DeltaAF=\muAF\Delta'$ [this includes AF and F phases; see Figs.~\ref{fig:spectra_CAF}(a), \ref{fig:spectra_CAF}(c), \ref{fig:spectra_CAF}(d)], the spectrum~(\ref{spectr_caf_zz}) is given by
\begin{equation}
E^{\xi\kappa}_{n\pm}(k)=
\pm\biggl|\sqrt{\mu'^2+\muAF^2}+\kappa\sqrt{\Delta'^2+\DeltaAF^2+\epsilon_0^2\lambda_n^\xi(k)}\biggr|.
\end{equation}
A pair of counterpropagating gapless edge states is present at each edge if $\sqrt{\mu'^2+\muAF^2}>\sqrt{\Delta'^2+\DeltaAF^2}$; otherwise the gap in the spectrum is equal to
\begin{equation}
E_{\rm gap}=2\biggl(\sqrt{\Delta'^2+\DeltaAF^2}-\sqrt{\mu'^2+\muAF^2}\biggr).
\end{equation}
Note that the gapless states  in the F phase with $|\mu'|<|\Delta'|$ are located between the two valleys~\cite{Kane2005PRL} and are not captured by the Dirac model~\cite{Gusynin2008PRB,Gusynin2009PRB,DeMartino2011PRB}.

In the absence of valley-odd potentials [$\muAF=\Delta'=0$; see Fig.~\ref{fig:spectra_CAF}(b)], the spectrum~(\ref{spectr_caf_zz}) is given by
\begin{equation}
E^{\xi\kappa}_{n\pm}(k)=
\pm\sqrt{\DeltaAF^2+\Bigl(|\mu'|+\kappa\epsilon_0\sqrt{\lambda_n^\xi(k)}\Bigr)^2},
\end{equation}
and the ratio of the edge gap $E_{\rm gap}=2|\DeltaAF|$ to the bulk gap $E_{0+}-E_{0-}=2\sqrt{\DeltaAF^2+\mu'^2}$ changes from unity in the AF phase ($\mu'=0$) to zero in the F phase ($\DeltaAF=0$). The behavior of edge state spectrum in this case qualitatively agrees with the recent numerical self-consistent Hartree-Fock study~\cite{Lado2014PRB}, where the modification of the order parameter at the boundary was taken into account.

In the case of a finite NNN hopping parameter, the boundary condition at $y=0$,
\begin{equation}
\sigma_0\otimes[\tau_0+\xi(\tau_3\cos\vartheta-\tau_1\sin\vartheta)]\Psi_{K_\xi}(-kl)=0,
\label{CAF_zigzag_BC_y0_mod}
\end{equation}
gives the dispersion equation
\begin{equation}
\begin{split}
&g^+u^{(-)}_{++}+\epsilon_0^2(t'/t)^{2\xi}u^{(-)}_{--}
+\xi(t'/t)^\xi\epsilon_0E\biggl\{u^{(-)}_{+-}+u^{(-)}_{-+} \\
&\;+\frac{4\bigl[\mu'(\Delta'+\mu')+\muAF(\DeltaAF+\muAF)\bigr](u^{(-)}_{+-}-u^{(-)}_{-+})}
{\epsilon_0^2(\lambda_+-\lambda_-)}\biggr\}=0,
\label{disp_eqn_zz_caf_t'}
\end{split}
\end{equation}
where we introduced
\begin{equation}
u^{(\pm)}_{\alpha\beta}=U\bigl(\alpha\tfrac12-\lambda_+,\pm\sqrt2kl\bigr)U\bigl(\beta\tfrac12-\lambda_-,\pm\sqrt2kl\bigr).
\end{equation}
For the edge $y=W$, the dispersion equation is obtained from~(\ref{disp_eqn_zz_caf_t'}) by replacing $k\to W/l^2-k$ and exchanging the valleys. Similarly to the CDW phase, the edge state branch connecting the two valleys becomes dispersive at finite $t'$ [Fig.~\ref{fig:spectra_CAF}(e)] and makes the spectrum gapless provided that $|t'|$ exceeds the LLL splitting $2E_{0+}$ with $E_{0+}$ given in Eq.~(\ref{LLL_CAF}).

In the case of an armchair ribbon, the boundary condition~(\ref{ac_bc}) at $x=x_0$ can be written as
\begin{equation}
\bigl[1+(\taut_2\cos\theta_0-\taut_1\sin\theta_0)\otimes\sigma_0\otimes\tau_2\bigr]
\begin{bmatrix}
\widetilde\Psi_{K_+} \\
\widetilde\Psi_{K_-}
\end{bmatrix}
_{x=x_0}=0,
\end{equation}
or, using Eq.~(\ref{zigzag_to_armchair}),
\begin{equation}
(1-\taut_2\otimes\sigma_0\otimes\tau_1)
\begin{bmatrix}
\Psi_{K_+} \\
e^{-i\theta_0}\Psi_{K_-}
\end{bmatrix}
_{\eta=kl+x_0/l}=0.
\end{equation}
Substituting the solution~(\ref{gen_sol_CAF}) into this equation with $x_0=0$ leads to the dispersion equation
\begin{equation}
\begin{split}
&g^+u^{(+)}_{++}-\epsilon_0^2u^{(+)}_{--}\pm
\biggl\{\epsilon_0(\mu'+\Delta')(u^{(+)}_{+-}+u^{(+)}_{-+}) \\
&\quad+\frac{4\bigl[\mu'(E^2-\DeltaAF^2)+\DeltaAF\muAF(\Delta'-\mu')+\Delta'\muAF^2\bigr]}
{\epsilon_0(\lambda_+-\lambda_-)}  \\
&\quad\times(u^{(+)}_{+-}-u^{(+)}_{-+})\biggr\}=0.
\label{disp_eq_ac_caf}
\end{split}
\end{equation}
For the edge $x=W$, the replacement $k\to-W/l^2-k$ has to be made in the above equation.

In the absence of the valley-odd potentials ($\muAF=\Delta'=0$) the dispersion equation~(\ref{disp_eq_ac_caf}) reduces to two identical equations~(\ref{disp_eqn_armchair_CDW_simpl}) for $\lambda=\lambda_\pm$. Therefore, the spectrum in this case is given by
\begin{equation}
E_{n\pm}^\kappa(k)=\pm\sqrt{\DeltaAF^2+\Bigl(|\mu'|+\kappa\epsilon_0\sqrt{\widetilde\lambda_n(k)}\Bigr)^2}, \qquad
\kappa=\pm.
\end{equation}
The edge gap, corresponding to the minimum value of the lowest positive energy branch $E_{0+}^{-}(k)$, is equal to $2|\DeltaAF|$~\cite{Kharitonov2012PRBa}. We find numerically that at nonzero $\muAF$ and $\Delta'$ ($|\muAF|,|\DeltaAF|\ll\epsilon_0$), the lowest branches of the spectrum have qualitatively similar behavior. The edge gap is approximately equal to $E_{\rm gap}\simeq2|\DeltaAF+\muAF|$ and is almost unaffected by the finite NNN hopping parameter (provided $|t'/t|\ll1$). The ratio of the edge gap to the bulk gap
\begin{equation}
\frac{E_{\rm gap}}{E_{0+}-E_{0-}}\simeq\biggl[1+\biggl(\frac{\mu'+\Delta'}{\muAF+\DeltaAF}\biggr)^2\biggr]^{-1/2}
\end{equation}
changes from unity in the AF phase to zero in the F phase [Figs.~\ref{fig:spectra_CAF}(f)--\ref{fig:spectra_CAF}(h)], in agreement with the previous theoretical results~\cite{Kharitonov2012PRBa,Lado2014PRB} and experiment~\cite{Young2014N}.

\section{Conclusion}
\label{secIV}

In this paper, we studied the edge state spectrum of the $\nu=0$ quantum Hall state in monolayer graphene in the CDW, KD, AF, CAF, and F phases. The main result is establishing the criterion for the existence of gapless current-carrying excitations in each phase, which provides the concrete theoretical predictions from the mean-field model with the homogeneous symmetry-breaking terms in the cases of ideal zigzag and armchair edges.

Our analysis shows that the existence of gapless edge states depends on the edge type, and the difference between the spectra of zigzag and armchair ribbons is even more profound in the case of a finite NNN hopping term.

For a ribbon with armchair edges, the influence of the NNN hopping parameter and the ratio of symmetry-breaking terms (chemical potentials and mass gaps) on the spectrum is negligible for all phases. In the CDW and AF phases, the band gap is equal to the bulk LLL splitting, in agreement with the previous studies~\cite{Gusynin2008PRB,Gusynin2009PRB,Jung2009PRB}. For the transition from the CAF to F phase, we obtain the closing of the edge gap, which is consistent with the earlier theoretical results~\cite{Kharitonov2012PRBa,Lado2014PRB} and the recent experiment~\cite{Young2014N}. In the KD phase, the spectrum is generically gapped but the edge gap closes at a certain valley isospin angle of the KD order parameter.

In the case of zigzag edges, the band gap is strongly affected by the finite NNN hopping parameter. At $t'=0$, the spectrum is gapped in the KD phase and gapless in the F phase. For the CDW and AF phases, the gapless edge states exist if the chemical-potential-like symmetry breaking terms exceed the corresponding mass gaps, in agreement with Refs.~\cite{Gusynin2008PRB,Gusynin2009PRB}. In the CAF phase, the band gap can vary between zero and the size of the bulk LLL splitting, depending on the ratios between four different symmetry-breaking terms. At a finite NNN hopping parameter larger than the bulk LLL splitting, the band gap is closed in all considered phases, except the KD one, due to the deformation of the edge state branch connecting the two valleys; for the KD phase, the edge gap becomes approximately equal to the half of the bulk gap. 
It is notable that the KD phase is the only state which can have the gapped spectrum at such a large value of NNN hopping for both edge types (as was already indicated in Sec. \ref{secIII}, the experimental value $|t'|\simeq 0.3$~eV~\cite{Kretinin2013PRB} is indeed large).

As shown in Ref.~\cite{Akhmerov2008PRB}, mixed armchair/zigzag edges with the intermediate orientation are generally described within the Dirac model by the zigzag-like boundary condition whereas the number of dispersionless edge states is determined by the percentage of zigzag edge segments or, equivalently, by the momentum separation of the $K_\pm$ points projected along the ribbon. This suggests that results obtained here for the zigzag case should hold in general for a mixed edge with the only difference being a reduced bandwidth of the intervalley edge state branch (given by $|t'|$ for a purely zigzag boundary). Therefore, in the cases when the spectrum is gapless due to this zigzag edge state branch, one can expect the gap opening at some critical deviation from the zigzag direction, when the edge state bandwidth becomes smaller than the bulk LLL splitting.

Our results for the case of armchair edges support the currently accepted CAF-F scenario~\cite{Kharitonov2012PRBa,Young2014N,Lado2014PRB,Murthy2014} of the observed gradual insulator-metal transition in the tilted magnetic field~\cite{Young2014N}. For the zigzag edges and finite NNN hopping, however, we find that CAF phase has gapless edge excitations. Whether these excitations indeed lead to a conducting state or they are modified substantially beyond the present model is an important question. As a first step, it would be reasonable to take into account the variation of the order parameter near the edges~\cite{Fertig2006PRL,*Shimshoni2009PRL,Jung2009PRB,Lado2014PRB,Murthy2014}.
This issue will be considered elsewhere.

\begin{acknowledgments}
This work was supported by the Natural Sciences and Engineering Research Council of Canada and by the Ontario Graduate Scholarship program.
\end{acknowledgments}

\section*{Appendix: Tight-binding Hamiltonian for graphene ribbon}

\renewcommand{\theequation}{A\arabic{equation}}
\setcounter{equation}{0}
\renewcommand*{\thesubsection}{\arabic{subsection}}

\subsection{Zigzag ribbon: CDW and CAF phases}

For a zigzag ribbon, the free part of the tight-binding Hamiltonian (including the NNN hopping terms) can be written as
\begin{equation}
\begin{split}
\mathcal H_0+\mathcal H'=-\!\!\!\int\limits_{-\pi/a}^{\pi/a}\!\!&\frac{dk}{2\pi}\,\chi^\dag(k)
\sigma_0\otimes\biggl(t
\begin{bmatrix}
0 & M_1 \\
M_1^T & 0
\end{bmatrix}
\\
&+2t'\re\begin{bmatrix}
M_2^+ & 0 \\
0 & M_2^-
\end{bmatrix}
\biggr)
\chi(k),
\end{split}
\end{equation}
where the $4N$ components of the vectors
\begin{equation}
\chi(k)=
\begin{bmatrix}
\chi^+_A(k) \\
\chi^+_B(k) \\
\chi^-_A(k) \\
\chi^-_B(k)
\end{bmatrix}
,\qquad
\chi^s_X(k)=
\begin{bmatrix}
c_{Xs1}(k) \\
c_{Xs2}(k) \\
\vdots \\
c_{XsN}(k)
\end{bmatrix}
,
\label{eq:chi(k)}
\end{equation}
are the Fourier-transformed in the $x$ direction lattice fermion operators,
\begin{equation}
\begin{bmatrix}
a_{jxs} \\
b_{jxs}
\end{bmatrix}
=\sqrt{a}\int\limits_{-\pi/a}^{\pi/a}\frac{dk}{2\pi}e^{ikx}
\begin{bmatrix}
c_{Asj}(k) \\
c_{Bsj}(k) \\
\end{bmatrix}
,\quad j=1,\dots,N.
\end{equation}
The symmetry-breaking terms~(\ref{Omega})--(\ref{Lambda}) are given by
\begin{align}
\Omega^\pm_\alpha&=\int\limits_{-\pi/a}^{\pi/a}\!\!\frac{dk}{2\pi}\,\chi^\dag(k)
\sigma_\alpha\otimes
\begin{bmatrix}
\mathbb1_N & 0 \\
0 & \pm\mathbb1_N
\end{bmatrix}
\chi(k),
\\
\Lambda^\pm_\alpha&=\int\limits_{-\pi/a}^{\pi/a}\!\!\frac{dk}{2\pi}\,\chi^\dag(k)
\sigma_\alpha\otimes\biggl(\frac2{3\sqrt3}\im
\begin{bmatrix}
M_2^+ & 0 \\
0 & \pm M_2^-
\end{bmatrix}
\biggr)
\chi(k).
\end{align}
The matrix elements of $M_1$ and $M_2^\pm$ are expressed as
\begin{equation}
\begin{split}
[M_1]_{jj'}={}&\delta_{j',j+1}+2\delta_{jj'}\cos(k_ja/2) \\
[M_2^\pm]_{jj'}={}&\delta_{jj'}e^{ik^\pm_ja} +(\delta_{j',j+1}+\delta_{j',j-1})e^{-i(k^\pm_j+k^\pm_{j'})a/4},
\end{split}
\end{equation}
where we introduced
\begin{equation}
k_j=k-\frac{\pi\phi}a\biggl(2j-\frac13\biggr), \qquad
k_j^\pm=k_j\mp\frac{\pi\phi}{3a}.
\end{equation}
Here $\phi=\sqrt3a^2/(4\pi l^2)$ is the magnetic flux through a hexagonal unit cell in units of the magnetic flux quantum.

\subsection{Zigzag ribbon: KD phase}

The Kekul\'e order term~(\ref{KD_term}) triples the number of nonequivalent atoms in the zigzag direction (Fig.~\ref{fig:lattice}), and the full mean-field Hamiltonian can be written as
\begin{equation}
\mathcal H_0+\mathcal H'+\mathcal H_{\rm KD}=\sum_{s=\pm}\int\limits_{-\frac\pi{3a}}^{\frac\pi{3a}}\frac{dk}{2\pi}
\chi^\dag_s(k)
H_s(k)\chi_s(k),
\end{equation}
where $\chi_s(k)$ is the $6N$-component vector
\begin{equation}
\chi_s(k)=
\begin{bmatrix}
\chi^s_{A_1}(k) \\
\chi^s_{A_2}(k) \\
\chi^s_{A_3}(k) \\
\chi^s_{B_1}(k) \\
\chi^s_{B_2}(k) \\
\chi^s_{B_3}(k)
\end{bmatrix}
,\qquad
\chi_X^s(k)=
\begin{bmatrix}
c_{Xs1}(k) \\
c_{Xs2}(k) \\
\vdots \\
c_{XsN}(k)
\end{bmatrix}
,
\end{equation}
and the blocks of the matrix
\begin{equation}
H_s(k)=
\begin{bmatrix}
Y_+ & X \\
X^\dag & Y_-
\end{bmatrix}
+s\mu_{\rm Z}\mathbb1_{6N},
\end{equation}
are given by
\begin{align}
X&=
\begin{bmatrix}
\beta_0M_3 & \beta_1M_4 & \beta_2M_4^\dag \\
\beta_1M_4^\dag & \beta_2M_3 & \beta_0M_4 \\
\beta_2M_4 & \beta_0M_4^\dag & \beta_1M_3
\end{bmatrix}
, \\
Y_\pm&=-t'
\begin{bmatrix}
0 & (M_2^\pm)^\dag & M_2^\pm \\
M_2^\pm & 0 & (M_2^\pm)^\dag \\
(M_2^\pm)^\dag & M_2^\pm & 0
\end{bmatrix}
.
\end{align}
The matrix elements of $M_3$ and $M_4$ are
\begin{equation}
\begin{split}
[M_3]_{jj'}&=\delta_{j',j+1}, \\
[M_4]_{jj'}&=\delta_{jj'}e^{ik_ja/2},
\end{split}
\end{equation}
and we introduced
\begin{equation}
\beta_j\equiv-t+\frac23\Delta\cos\Bigl(\theta-\frac{2\pi j}3\Bigr)+\frac{2i}3\mu\sin\Bigl(\theta-\frac{2\pi j}3\Bigr).
\end{equation}

\subsection{Armchair ribbon}

For an armchair ribbon, the free part of the Hamiltonian (including the NNN hopping terms) reads
\begin{equation}
\begin{split}
\mathcal H_0+\mathcal H'=-\int\limits_{-\frac\pi{\sqrt3a}}^{\frac\pi{\sqrt3a}}&\frac{dk}{2\pi}\,\chi^\dag(k)
\sigma_0\otimes\biggl(t
\begin{bmatrix}
0 & M_5 \\
M_5^\dag & 0
\end{bmatrix}
\\
&+t'
\begin{bmatrix}
M_6^+ & 0 \\
0 & M_6^+
\end{bmatrix}
\biggr)
\chi(k),
\end{split}
\end{equation}
where $\chi(k)$ is defined in Eq.~(\ref{eq:chi(k)}) with
\begin{equation}
\begin{bmatrix}
a_{jys} \\
b_{jys}
\end{bmatrix}
=\bigl(\sqrt3a\bigr)^{\frac12}\!\!\!\int\limits_{-\frac\pi{\sqrt3a}}^{\frac\pi{\sqrt3a}}\!\frac{dk}{2\pi}e^{iky}
\begin{bmatrix}
c_{Asj}(k) \\
c_{Bsj}(k) \\
\end{bmatrix}
,
\quad
j=1,\dots,N.
\end{equation}
The symmetry-breaking terms~(\ref{Omega})--(\ref{Lambda}) are given by
\begin{equation}
\Omega^\pm_\alpha=\int\limits_{-\frac\pi{\sqrt3a}}^{\frac\pi{\sqrt3a}}\frac{dk}{2\pi}\,\chi^\dag(k)
\sigma_\alpha\otimes
\begin{bmatrix}
\mathbb1_N & 0 \\
0 & \pm\mathbb1_N
\end{bmatrix}
\chi(k),
\end{equation}
\begin{equation}
\Lambda^\pm_\alpha=3^{-\frac32}i\int\limits_{-\frac\pi{\sqrt3a}}^{\frac\pi{\sqrt3a}}\frac{dk}{2\pi}\,\chi^\dag(k)
\sigma_\alpha\otimes
\begin{bmatrix}
M_6^- & 0 \\
0 & \pm M_6^-
\end{bmatrix}
\chi(k),
\end{equation}
\begin{equation}
\mathcal H_{\rm KD}=\int\limits_{-\frac\pi{\sqrt3a}}^{\frac\pi{\sqrt3a}}\frac{dk}{2\pi}\,\chi^\dag(k)
\sigma_0\otimes
\begin{bmatrix}
0 & M_7 \\
M_7^\dag & 0
\end{bmatrix}
\chi(k)+\mathcal H_{\rm Z}.
\end{equation}
The matrix elements of $M_5$, $M_6^\pm$, and $M_7$ are
\begin{equation}
\begin{split}
[M_5]_{jj'}&=\delta_{j'j}e^{\frac{i\widetilde k_ja}{\sqrt3}}
+(\delta_{j',j+1}+\delta_{j',j-1})e^{-\frac{i(\widetilde k_j+\widetilde k_{j'})a}{4\sqrt3}}, \\
[M_6^\pm]_{jj'}&=\delta_{j',j-2}\pm\delta_{j',j+2} \\
&\hspace{-7mm}+2(\delta_{j',j+1}\pm\delta_{j',j-1})\cos\Bigl(\bigl(\widetilde k_{j}+\widetilde k_{j'}\bigr)a\sqrt3/4\Bigr), \\
[M_7]_{jj'}&=(\beta_{j+j'}+t)[M_5]_{jj'},
\end{split}
\end{equation}
where $\widetilde k_j=k+2\pi\phi j/(\sqrt3a)$.

\bibliography{bibliography}

\end{document}